\documentclass[onecolumn,showpacs,nobibnotes,nofootinbib,12pt,aps,prd,showpacs,notitlepage,nofootinbib,preprintnumbers,amsmath,amssymb]{article}

\pdfoutput=1

\usepackage{jheppub}

\usepackage{amsmath,amssymb}
\usepackage{wasysym}
\usepackage{slashed}
\usepackage{graphicx}
\usepackage{pdfcomment}

\usepackage{graphics,graphicx}
\usepackage{dcolumn}
\usepackage{bm}
\usepackage{epsfig}
\usepackage{hyperref} 
\usepackage{mathbbol}
\usepackage{epstopdf}
\usepackage{simplewick}
\usepackage[utf8]{inputenc} 
\usepackage{slashed}

\def\eq#1{{Eq.~(\ref{#1})}}
\def\fig#1{{Fig.~\ref{#1}}}
\newcommand{\ben}{\begin{eqnarray*}}
\newcommand{\een}{\end{eqnarray*}}
\newcommand{\un}[1]{\underline{#1}}

\newcommand{\pd}{\partial}

\newcommand{\stackeven}[2]{{{}_{\displaystyle{#1}}\atop\displaystyle{#2}}}
\newcommand{\lsim}{\stackeven{<}{\sim}}

\newcommand{\as}{\alpha_s}

\newcommand{\dhd}{{\textstyle d}
\lower.03ex\hbox{\kern-0.38em$^{\scriptstyle-}$}\kern-0.05em{}}
\newcommand{\dbar}{{\textstyle \delta}
\lower.03ex\hbox{\kern-0.38em$^{\scriptstyle-}$}\kern-0.05em{}}

\title{{\boldmath Time-Dependent Observables in Heavy Ion Collisions
    II: in Search of Pressure Isotropization in the $\varphi^4$
    Theory}}

\author{Yuri V. Kovchegov,}
\author{Bin Wu}

\affiliation{Department of Physics, The Ohio State University, Columbus, OH 43210, USA}

\emailAdd{kovchegov.1@osu.edu}
\emailAdd{bin.wu.phys@gmail.com}

\abstract{To understand the dynamics of thermalization in heavy ion
  collisions in the perturbative framework it is essential to first
  find corrections to the free-streaming classical gluon fields of the
  McLerran--Venugopalan model. The corrections that lead to deviations
  from free streaming (and that dominate at late proper time) would
  provide evidence for the onset of isotropization (and, possibly,
  thermalization) of the produced medium. To find such corrections we
  calculate the late-time two-point Green function and the
  energy-momentum tensor due to a single $2 \to 2$ scattering process
  involving two classical fields. To make the calculation tractable we
  employ the scalar $\varphi^4$ theory instead of QCD. We compare our
  exact diagrammatic results for these quantities to those in kinetic
  theory and find disagreement between the two. The disagreement is in
  the dependence on the proper time $\tau$ and, for the case of the
  two-point function, is also in the dependence on the space-time
  rapidity $\eta$: the exact diagrammatic calculation is, in fact,
  consistent with the free streaming scenario. Kinetic theory predicts
  a build-up of longitudinal pressure, which, however, is not observed
  in the exact calculation. We conclude that we find no evidence for
  the beginning of the transition from the free-streaming classical
  fields to the kinetic theory description of the produced matter
  after a single $2 \to 2$ rescattering.}

\begin{document}

\maketitle
\flushbottom


\section{Introduction}
\label{sec:intro}

The problem of isotropization and thermalization of the medium
produced in ultra-relativistic heavy ion collisions is arguably the
central theoretical problem in the field since it addresses the
fundamental question of whether and how quark-gluon plasma (QGP) is
formed in these collisions. Despite a number of theoretical efforts,
the solution of this problem still remains elusive. Thermalization
appears to be easier to tackle at strong (`t Hooft) coupling in the
framework of the anti-de Sitter/Conformal Field Theory (AdS/CFT)
correspondence \cite{Maldacena:1997re,Witten:1998qj}: there it is
possible to show that a collision of two shock waves results in the
black hole formation in the AdS$_5$ bulk, corresponding to a thermal
medium being formed at the boundary. This was demonstrated
analytically (but indirectly) using the trapped surface analysis
\cite{Gubser:2008pc,Lin:2009pn,Kovchegov:2009du} and was directly
observed in a numerical solution of Einstein equations
\cite{Chesler:2009cy,Chesler:2010bi}. The weakness of the approach
based on AdS/CFT correspondence is that the duality is for ${\cal
  N}=4$ super-Yang-Mills theory and not for quantum chromodynamics
(QCD). Nevertheless, a consensus exists in the community that
thermalization of the medium produced in high-energy collisions in
strongly-coupled field theories is very likely to take place and to
happen on a very short time scale.

The efforts to tackle the isotropization and thermalization problems
at weak coupling have not achieved such a universal consensus. The
initial break-through in the theoretical understanding of
thermalization in QCD at weak coupling was the so-called `bottom-up
thermalization' proposal \cite{Baier:2000sb}. In this scenario, the
quark-gluon system that begins in an initial state due to saturated
gluon fields created in nuclear collisions (dominated by the classical
gluon fields of the McLerran-Venugopalan (MV) model
\cite{McLerran:1993ni,McLerran:1993ka,McLerran:1994vd,Krasnitz:1998ns,Krasnitz:1999wc,Krasnitz:2003nv,Lappi:2003bi})
progresses to a thermalized isotropic medium due to $2 \to 2$, $2 \to
3$ and $3 \to 2$ rescatterings. However, the qualitative arguments
presented in \cite{Baier:2000sb} have never been verified by explicit
diagrammatic calculations. Moreover, in
\cite{Arnold:2003rq,Arnold:2004ih,Arnold:2004ti} it was pointed out
that the bottom-up thermalization scenario may be invalidated by the
occurrence of plasma instabilities, which could be present due to the
momentum-space anisotropy of the initial non-equilibrium gluonic
medium. Numerical simulations of these instabilities appear to
demonstrate that the growth of instabilities is stopped due to the
non-Abelian nature of strong interactions \cite{Rebhan:2004ur,
  Romatschke:2006nk}, possibly reinstating the original `bottom-up'
scenario. Alternative approaches \cite{Berges:2013fga} apply classical
Yang-Mills dynamics to confirm the parametric scaling of the
observables predicted in the first stage of the `bottom-up'
thermalization \cite{Baier:2000sb}, and possibly leading to eventual
thermalization of the medium \cite{Gelis:2013rba}: however, the
resulting classical dynamics appears to be non-renormalizable
\cite{Epelbaum:2014yja,Epelbaum:2014mfa}.

Yet another weakly-coupled approach to thermalization is based on
Boltzmann equation. The applicability of Boltzmann equation to
description of the medium produced in late stages of heavy ion
collisions was argued in \cite{Mueller:2002gd,Arnold:2002zm} (with the
Vlasov--Boltzmann equation used in the instability studies mentioned
above). Boltzmann equation dynamics appears to lead to thermalization
of the produced medium \cite{Kurkela:2015qoa}, confirming the
bottom-up thermalization scenario. However, the existing literature
lacks a side-by-side comparison of Boltzmann equation with the
explicit diagram calculation for heavy ion collisions: such a
comparison is needed to either validate the applicability of the
Boltzmann equation to heavy ion collisions or to prove
otherwise. Performing such a cross-check is the main goal of the
present work. In \cite{Mueller:2002gd} a correspondence was
established between the Boltzmann equation containing only the
order-$f^3$ part of the collision term and the classical gluon fields
at late times (and, hence, the late-time limit of the Feynman diagrams
describing the collision in the classical approximation). However, the
correspondence was never checked for the Boltzmann equation with the
order-$f^2$ part of the collision term and the diagrams describing
some correction to the classical fields of the MV model. This will be
performed below.

In more general terms, many perturbative thermalization scenarios
assume that the classical gluon fields of the MV model become
sub-leading at late proper times in the collisions, being superseded
by fields produced by some other dynamics, for instance due to the
Boltzmann equation. However, no explicit calculation of Feynman
diagrams exists in the literature which starts with the actual
collision of two large nuclei, identifies a particular diagrammatic
correction to the classical gluon fields and shows explicitly that
such a correction becomes dominant at late times. The absence of such
calculations is probably attributable to their complexity. However, an
approach like this would have been natural in the saturation/Color
Glass Condensate (CGC) framework
\cite{Gribov:1984tu,Iancu:2003xm,Weigert:2005us,Jalilian-Marian:2005jf,Gelis:2010nm,Albacete:2014fwa,KovchegovLevin}. There
(and elsewhere in perturbative calculations in field theory) one
usually starts with the tree-level leading-order contribution, which
is often classical. The leading-order contribution receives
corrections due to quantum fluctuations, leading parts of which may be
resummed using evolution equations. This program has been carried out
for the case of deep inelastic scattering (DIS) at small Bjorken $x$,
where the leading order contribution to unpolarized DIS structure
functions is given by the Glauber--Mueller quasi-classical multiple
rescatterings \cite{Mueller:1989st}, and the quantum corrections
resumming logarithms of $1/x$ are included via the Balitsky--Kovchegov
(BK)
\cite{Balitsky:1995ub,Balitsky:1998ya,Kovchegov:1999yj,Kovchegov:1999ua}
and Jalilian-Marian--Iancu--McLerran--Weigert--Leonidov--Kovner
(JIMWLK) evolution equations
\cite{Jalilian-Marian:1997dw,Jalilian-Marian:1997gr,Iancu:2001ad,Iancu:2000hn}.

For heavy ion collisions analyzed in the saturation framework the
leading contribution to, say, the energy-momentum tensor of the
produced medium is given by the classical gluon field of the MV model
\cite{McLerran:1993ni,McLerran:1993ka,McLerran:1994vd,Krasnitz:1998ns,Krasnitz:1999wc,Krasnitz:2003nv,Lappi:2003bi}. This
is already a very difficult calculation, only possible to be fully
done numerically due to the complexity of the analytic attempts
\cite{Balitsky:2004rr,Chirilli:2015tea} (see
\cite{Kovner:1995ts,Kovner:1995ja,Kovchegov:1997ke,Kovchegov:1998bi,Dumitru:2001ux}
for perturbative results valid for proton-proton and proton-nucleus
collisions). The numerical calculations
\cite{Krasnitz:1998ns,Krasnitz:1999wc,Krasnitz:2003nv,Lappi:2003bi}
indicate that the classical gluon fields lead to a free-streaming
medium, characterized by zero longitudinal pressure $P_L =0$ and the
energy density $\epsilon = 2 \, P_T \sim 1/\tau$ at late proper times
$\tau \gg 1/Q_s$. (Here $P_T$ and $P_L$ are the transverse and
longitudinal pressures at mid-rapidity, $\tau = \sqrt{t^2 - z^2}$ is
the proper time, and $Q_s$ is the classical gluon saturation scale.)
Quantum corrections to the classical energy-momentum tensor resumming
leading logarithms of $1/x$ were addressed in
\cite{Gelis:2008rw,Gelis:2008ad}, where it was argued that such
corrections can be resummed using the JIMWLK evolution equation for
the weight functionals of the color charge densities in the two
nuclei. The resulting gluon fields are still obtained by solving the
classical Yang-Mills equations, but now with the JIMWLK-modified
distribution of color sources. Therefore, such small-$x$ evolution
corrections still lead to a classical free-streaming energy-momentum
tensor and are not related to isotropization or thermalization of the
medium.

The question of whether perturbative quantum corrections to the
classical gluon fields which usher in isotropization and
thermalization exist still remains open. Over a decade ago, one of the
authors of this work tried looking for such corrections in
\cite{Kovchegov:2005ss} (see also \cite{Kovchegov:2005kn}). Having
failed to find them, he argued that such corrections do not exist as
long as one can define a gluon production cross section: hence, the
end state of any perturbative (weakly-coupled) dynamics in heavy ion
collisions was argued to always be a free-streaming bunch of particles
\cite{Kovchegov:2005ss,Kovchegov:2005kn}.

The arguments of \cite{Kovchegov:2005ss,Kovchegov:2005kn}
notwithstanding, potential candidates for the isotropization-inducing
quantum corrections are the $2\to 2$ rescatterings as resummed by
Boltzmann equation in the framework of kinetic theory. In the previous
part I of this paper duplex \cite{WuKovchegov} we showed that if one
starts with the Boltzmann distribution function $f^{(0)}$ for the
classical gluon fields of the MV model, and inserts it into the
order-$f^2$ part of the collision term of the Boltzmann equation,
solving the latter for a corrected distribution $f^{(1)}$, one indeed
does obtain isotropization corrections to the $P_L =0$, $\epsilon \sim
1/\tau$ free-streaming behavior of the classical gluon medium. The
remaining question is whether Boltzmann equation correctly represents
the Feynman diagrams it purports to sum. In \cite{WuKovchegov} we
review the derivation of Boltzmann equation that exists in the
literature, concentrating on the same case of a single $2\to 2$
rescattering of two classical gluon fields. The conclusion reached in
\cite{WuKovchegov} is that the underlying Feynman diagrams including
the $2\to 2$ rescattering lead either to results consistent with
Boltzmann equation prediction of to free-streaming depending on how
the late-time limit is taken. Namely, denote by $\tau_0$ the time of
the $2\to 2$ rescattering (assumed to be instantaneous in the
derivation, which is valid at late times only when the gradient
expansion becomes possible) and by $\tau$ the time in the argument of
$f$ (that is, the time when we measure the particle in question). For
the Boltzmann equation to be valid, the particles (gluons) must
approximately go on mass-shell both in the time after they are
produced in a collision but before they rescatter and in the time
after they rescatter but before they are detected. This means that
time intervals $\tau_0$ and $\tau - \tau_0$ should be sufficiently
long. While it is clear that for $\tau_0$ ``sufficiently long'' (on
the average) means $\tau_0 \gg 1/Q_s$, since $1/Q_s$ is the time it
takes for the classical gluon fields to go on mass shell, it is less
clear what ``sufficiently long'' means for $\tau - \tau_0$. In
\cite{WuKovchegov} we consider two options,
\begin{itemize}
\item[(i)] $\tau - \tau_0 \gg \frac{1}{Q_s}$, \hspace*{5mm} $\tau_0 \gg 1/Q_s$; 
\item[(ii)] $\tau - \tau_0 \gg \tau_0 \gg 1/Q_s$
\end{itemize}
and show that the ordering (i) yields results consistent with the
Boltzmann equation, while the ordering (ii) leads to free streaming
and is not consistent with the Boltzmann equation. Unfortunately the
calculation performed in \cite{WuKovchegov} (along with the earlier
arguments in favor of Boltzmann equation) was too coarse to tell us
whether the ordering (i) or the ordering (ii) follows from the full
Feynman diagram calculation. It is the goal of the present paper to
resolve this ambiguity, at least in the framework of the $\varphi^4$
theory that we use for simplicity instead of QCD.

Below we will calculate the Feynman diagrams contributing to the $2\to
2$ rescattering of two classical fields using the Schwinger-Keldysh
formalism. As we have already mentioned, we will use the scalar
$\varphi^4$ theory coupled to an external current \cite{Gelis:2006yv}
for simplicity. Without going into detail of the scalar particle
production (though one could think of the Higgs production via gluon
fusion), we assume that two scalar particles were produced in the
collision with their distribution given by the two-point correlation
functions very similar to that for the classical gluon fields in the
saturation/CGC physics. (These two particles do not have to be on mass
shell.) The setup of the problem is presented in Sec.~\ref{sec:iso}
below. The particles rescatter via the $2\to 2$ process, which is
simpler in the $\varphi^4$ theory than in QCD: for instance, this
interaction is truly instantaneous in the $\varphi^4$ theory. The
two-point coordinate-space correlation function $G (x_1, x_2)$
resulting from the rescattering is calculated in
Sec.~\ref{sec:x1x2}. A calculation of the mixed-representation Green
function $G(X,P)$ (that is commonly used in derivations of Boltzmann
equation) resulting from the same $2\to 2$ rescattering process is
presented in Sec.~\ref{sec:XP}. (Sec.~\ref{sec:x1x2} also contains a
calculation of the corresponding energy-momentum tensor.) Both Green
function calculations lead to the result consistent with free
streaming and hence with the case (ii). Therefore, we see no evidence
supporting the use of Boltzmann equation in describing the
(perturbative) dynamics of the medium produced in heavy ion
collisions. Our conclusions are summarized in Sec.~\ref{sec:disc}.


\section{Isotropization problem for the $\varphi^4$ theory}
\label{sec:iso}

\subsection{Classical two-point correlation function}

The two-point $22$ correlation function due to the lowest-order
classical gluon fields in the MV model was calculated in
\cite{WuKovchegov} using the $A^+=0$ light-cone gauge. The result is
\begin{align}\label{Gcl}
  G_{22}^{a\mu,b\nu}(k, k')\equiv & \langle A^{a \mu} (k) \, A^{b \nu}
  (k') \rangle =
  (2\pi)^2\delta(\underline{k} + \underline{k}') \frac{i}{k^2 + i \epsilon k^0} \frac{i}{k^{\, \prime \, 2} + i \epsilon k^{\, \prime \, 0}} \nonumber\\
  & \times \left( -\frac{16\pi^2\alpha_s^3\, \delta^{ab}}{N_c}\right)
  \left(\frac{A}{S_\perp}\right)^2 \frac{1}{(\underline{k}^2)^2} \,
  \ln\frac{\underline{k}^2}{\Lambda^2} \,
  \sum\limits_{\lambda=\pm}\epsilon_\lambda^\mu(k)
  \epsilon_{\lambda}^{*\nu}(-k')
\end{align}
with the polarization vector
\begin{align}
  \epsilon^\mu_\lambda(k)= \left( 0,
    \frac{\underline{\epsilon}_\lambda \cdot \underline{k}}{k^+},
    \underline{\epsilon}_\lambda \right) \qquad \text{and} \qquad
  \sum\limits_{\lambda=\pm} \epsilon_\lambda^i
  \epsilon^{*j}_\lambda=\delta^{ij}.
\end{align}
Here we assume the collision of two identical large nuclei with atomic
numbers $A_1 = A_2 = A$, each of them shaped as a
longitudinally-oriented cylinder with a very large cross-sectional
area $S_\perp$. Underlined variables denote two-dimensional vectors in
transverse plane, $\un{v} = (v^1, v^2)$, while the light-cone
variables are $v^\pm = (v^0 \pm v^3)/\sqrt{2}$ with $x^3 = z$ the
collision axis. The contributing diagrams for the correlator
\eqref{Gcl} are shown in Fig.~3 of \cite{WuKovchegov} and are
comprised of two sets of lowest-order gluon production diagrams
(cf. \cite{Kovner:1995ts,Kovner:1995ja,Kovchegov:1997ke}).

\begin{figure}[htb]
\begin{center}
\includegraphics[width=0.25 \textwidth]{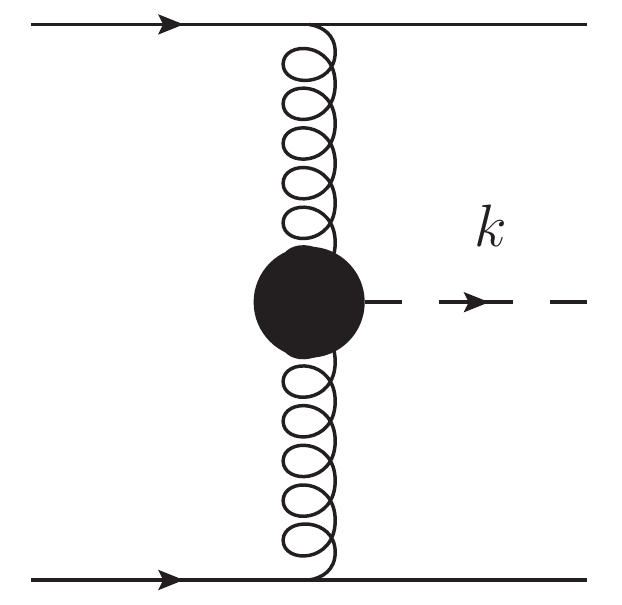}
\end{center}
\caption{Scalar particle production as envisioned here, with the solid
  lines denoting colliding quarks in the two nuclei and the dashed
  line denoting the scalar particle. The shaded circle may represent a
  quark loop, as would be the case in the Higgs production.}
\label{fig:scalar_prod}
\end{figure}

As described above, it would be very interesting and important to find
the perturbative correction to the correlator \eqref{Gcl} due to a
$2\to 2$ rescattering process involving two such Green
functions. However, full calculation of the order-$\as^2$ correction
to \eq{Gcl} involving two classical correlators (that is, a
calculation of the order-$\as^8 (A/S_\perp)^4$ correlator) appears to
be prohibitively complicated in QCD. Instead we will tackle a similar
problem in massless $\varphi^4$ theory. To do so we first have to
construct an analogue of the correlator \eqref{Gcl} in the massless
scalar theory. This is achieved by replacing the polarization sum in
\eq{Gcl} by $(-1)$ and writing the rest of the expression as
\begin{align}
  \label{eq:Gcl_sc}
  G^{LO}_{22} (k, k') \equiv \langle \varphi_{cl} (k) \, \varphi_{cl}
  (k') \rangle = -\frac{i}{k^2 + i \epsilon k^0} \, f(k_T) \, (2 \pi)^2
  \, \delta^2 (\un{k} + \un{k}') \, \frac{i}{k^{\, \prime \, 2} + i
    \epsilon k^{\, \prime \, 0}}
\end{align}
with $f(k_T)$ a function of the magnitude of the transverse momentum
$k_T = |\un{k}|$ which falls off rather fast at large $k_T$ and is
infrared (IR) finite due to saturation effects. The exact form of
$f(k_T)$ is not going to be important below. (This function is
proportional to the $k_T$ spectrum $dN/d^2 k_T dy$ of the produced
particles in the classical approximation \cite{Kovchegov:2005ss}.) The
rapidity-independent correlation function \eqref{eq:Gcl_sc} can not
result from a collision of particles taken entirely in the scalar
theory: instead, one can think of it as resulting from some
gluon+gluon$\to$scalar fusion process, with the two gluons coming from
the classical fields of the two colliding nuclei, as schematically
shown in \fig{fig:scalar_prod}, where the shaded circle represents an
effective gluon+gluon$\to$scalar vertex. Higgs production through
gluon fusion (via a top-quark loop) is one example of such a process
(though of course Higgs is massive, unlike the massless scalar
considered here). Let us stress one more time that the exact origin of
the correlation function \eqref{eq:Gcl_sc} is not important here: what
is important is that it carries the main features of the gluon
correlation function \eqref{Gcl}.

\begin{figure}[htb]
\begin{center}
\includegraphics[width=0.4 \textwidth]{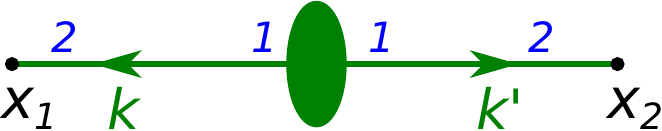}
\end{center}
\caption{ The classical correlation function from \eq{eq:Gcl_sc} or
  \eq{eq:Gcl_coord2} with the green oval denoting all the possible
  contributing Feynman diagrams (like the `square' of the one shown in
  \fig{fig:scalar_prod}). Solid lines from now on represent the scalar
  field. The indices $1$ and $2$ denote the type of the propagator in
  the Schwinger--Keldysh formalism.}
\label{fig:g22cl}
\end{figure}

Our notation for the correlation function \eqref{eq:Gcl_sc} is shown
in \fig{fig:g22cl}, where the Feynman diagrams contributing to the
correlator are summarily shown by a green oval.

In coordinate space the Green function \eqref{eq:Gcl_sc} is given by
the Fourier transform
\begin{align}
  \label{eq:Gcl_coord}
  G^{LO}_{22} (x_1, x_2) = \int \frac{d^4 k}{(2 \pi)^4} \, \frac{d^4
    k'}{(2 \pi)^4} \, e^{- i k \cdot x_1 - i k' \cdot x_2}
  \,G^{LO}_{22} (k, k').
\end{align}
Employing the integrals in the appendices of \cite{Kovchegov:2005ss}
we obtain
\begin{align}
  \label{eq:Gcl_coord2}
  G^{LO}_{22} (x_1, x_2) = \frac{1}{4} \int \frac{d^2 k_T}{(2
    \pi)^2} \,e^{i \un{k} \cdot \un{x}_{12}} \, f(k_T) \, J_0 (k_T \,
  \tau_1) \, J_0 (k_T \, \tau_2),
\end{align}
where $\un{x}_{12} = \un{x}_{1} - \un{x}_{2}$ and $\tau_1 = \sqrt{2
  x_1^+ x_1^-}$, $\tau_2 = \sqrt{2 x_2^+ x_2^-}$. Using the
large-argument asymptotics of Bessel functions we conclude that
\begin{align}
  \label{eq:Gcl_coord3}
  G^{LO}_{22} (x_1, x_2) \Bigg|_{\tau_1 = \tau_2 = \tau \gg
    \frac{1}{Q_s}} = \frac{1}{4 \pi \tau} \int \frac{d^2 k_T}{(2 \pi)^2}
  \,e^{i \un{k} \cdot \un{x}_{12}} \, \frac{f(k_T)}{k_T},
\end{align}
where we have averaged the cosine squared over time. (We have also
employed the fact that $f(k_T)$ is regulated by the saturation scale
$Q_s$ in the IR, such that the $k_T \approx 0$ region does not
contribute significantly to the integral.) 

Importantly the classical correlation function scales as
\begin{align}
  \label{eq:Gcl_coord4}
  G^{LO}_{22} (x_1, x_2) \Bigg|_{\tau_1 = \tau_2 = \tau \gg
    \frac{1}{Q_s}} \propto \frac{1}{\tau},
\end{align}
which is a tell-tale sign of free streaming. To understand this
better, let us calculate the energy-momentum tensor corresponding to
this classical dynamics. The correlation function
\eqref{eq:Gcl_coord2} is independent of the space-time rapidities
$\eta_1 = (1/2) \ln (x_1^+/x_1^-)$ and $\eta_2 = (1/2) \ln
(x_2^+/x_2^-)$. It is natural to conclude, and we will see shortly
that this is indeed the case, that the corresponding energy-momentum
tensor is also rapidity-independent. 

The most general energy-momentum tensor for a medium produced in a
high-energy collision of two large nuclei with the
rapidity-independent matter distribution can be parametrized in terms
of the energy density $\epsilon$ and transverse and longitudinal
pressures $P_T, P_L$ as
\begin{align}\label{tmngen}
  & T^{++} \, = \, [\epsilon (\tau) + P_L (\tau)] \, \left(
    \frac{x^+}{\tau}
  \right)^2, \nonumber ~\\
  & T^{--} \, = \, [\epsilon (\tau) + P_L (\tau)] \, \left(
    \frac{x^-}{\tau} \right)^2, \nonumber ~\\
  & T^{+-} \, = \,  [\epsilon (\tau) - P_L (\tau)] \, \frac{1}{2}, \nonumber ~\\
  & T^{ij} \, = \, \delta^{ij} \, P_T (\tau).
\end{align}
(We employ translational invariance in the transverse plane due to the
nuclei being very large.) At mid-rapidity ($z=0$) the tensor
(\ref{tmngen}) looks like
\begin{align}
  T^{\mu\nu} (z=0) = \left(
\begin{matrix}
\epsilon (\tau) & 0 & 0 & 0 \\
  0 & P_T (\tau) & 0 & 0 \\
  0 & 0 & P_T (\tau) & 0  \\
  0 & 0 & 0  & P_L (\tau) \\ 
\end{matrix}
\right)
\end{align}
in the $t,x,y,z$ coordinates. It is also useful to note that the
energy-momentum conservation
\begin{align}
  \partial_\mu T^{\mu\nu} \, = \, 0
\end{align}
yields
\begin{align}\label{bjhyd}
  \frac{d \epsilon}{d \tau} \, = \, - \frac{\epsilon + P_L}{\tau}.
\end{align}

The energy-momentum tensor in the massless $\varphi^4$ theory with the
Lagrangian density
\begin{align}
  \label{eq:L} {\cal L} = \frac{1}{2} \pd_\mu \varphi \, \pd^\mu
  \varphi - \frac{\lambda}{4!} \, \varphi^4
\end{align}
is given by 
\begin{align}
  \label{eq:tmnsc}
  T^{\mu\nu} = \langle \pd^\mu \varphi \, \pd^\nu \varphi - g^{\mu\nu}
  \, {\cal L} \rangle = \left\langle \pd^\mu \varphi \, \pd^\nu
    \varphi - g^{\mu\nu} \, \left[ \frac{1}{2} \pd_\alpha \varphi \,
      \pd^\alpha \varphi - \frac{\lambda}{4!} \, \varphi^4 \right]
  \right\rangle .
\end{align}
With the two-point correlation function decaying at late times as
shown in \eq{eq:Gcl_coord4}, it is natural to conclude that the
four-point correlation function decays twice as fast and is,
therefore, negligible in \eq{eq:tmnsc} taken at late times. We
therefore arrive at
\begin{align}
  \label{eq:tmncs_late}
  T^{\mu\nu}_{LO} \Bigg|_{\tau \gg \frac{1}{Q_s}} = \left[ \pd_1^\mu
    \, \pd_2^\nu - \frac{1}{2} g^{\mu\nu} \, \pd_{1 \, \alpha} \,
    \pd_2^\alpha \right] \, G^{LO}_{22} (x_1, x_2) \Bigg|_{x_1 = x_2,
    \ \tau \gg \frac{1}{Q_s}} \, .
\end{align}
Here the subscripts $1$ and $2$ denote derivatives with respect to
$x_1$ and $x_2$ respectively. Substituting the Green function from
\eq{eq:Gcl_coord2} we find (cf. \cite{Kovchegov:2005ss,Lappi:2006hq},
the $J_0 \leftrightarrow J_1$ difference in $p_T$ is inconsequential
at late times)
\begin{subequations} \label{eq:epp}
  \begin{align}
    & \epsilon = \frac{1}{8} \, \int \frac{d^2 k_T}{(2 \pi)^2} \,
    f(k_T) \, k_T^2 \, \left\{ \left[ J_0 (k_T \, \tau) \right]^2 +
      \left[ J_1 (k_T \, \tau) \right]^2 \right\} , \\
    & P_T = \frac{1}{8} \, \int \frac{d^2 k_T}{(2 \pi)^2} \, f(k_T)
    \, k_T^2 \, \left[ J_1 (k_T \, \tau) \right]^2 \\
    & P_L = \frac{1}{8} \, \int \frac{d^2 k_T}{(2 \pi)^2} \, f(k_T) \,
    k_T^2 \, \left\{ \left[ J_1 (k_T \, \tau) \right]^2 - \left[ J_0
        (k_T \, \tau) \right]^2 \right\}
  \end{align}
\end{subequations}
for the classical medium at $\tau Q_s \gg 1$. Applying the late-time
asymptotics to Eqs.~\eqref{eq:epp} we arrive at
\begin{align}
  \label{eq:free}
  \epsilon = 2 \, P_T \sim \frac{1}{\tau}, \ \ \ p_L =0.
\end{align}
This is a free-streaming anisotropic medium, with zero longitudinal
pressure and a non-zero transverse pressure. In the full MV model, the
classical gluon fields produced in a nuclear collision also lead to
the free-streaming asymptotics of \eq{eq:free}.

Note also that due to \eq{bjhyd} the $\epsilon \sim 1/\tau$ scaling
corresponds to $P_L =0$ and, hence, to free streaming. A more
isotropic medium with non-zero $P_L >0$ would have the energy density
$\epsilon$ falling off faster than $1/\tau$. The isotropization
problem in heavy ion collisions can be formulated as follows: can one
find (presumably quantum) corrections to the classical field
correlator \eqref{eq:Gcl_sc} modifying the free-streaming result
\eqref{eq:free} at late times in such a way as to generate a non-zero
$p_L >0$, or, equivalently, give us the net energy density $\epsilon$
that decreases faster than $1/\tau$?

\subsection{Boltzmann equation prediction for the two-point function
  after a single rescattering}

Kinetic theory appears to give a positive answer to the above
isotropization question. In \cite{WuKovchegov} we show that using the
classical particle distribution $f_{cl} = f^{(0)}$ resulting from the
correlator \eqref{Gcl}, inserting it in the collision term of the
Boltzmann equation, and solving the resulting equation for the new
distribution $f^{(1)}$ gives the following components of the
energy-momentum tensor:
\begin{subequations}\label{eq:results}
\begin{align}
  & \epsilon = \epsilon^{(0)} + \epsilon^{(1)} = \frac{A^{(0)} + \as^2 \, A^{(1)}}{\tau} - \frac{\as^2 \, B^{(1)}}{\tau} \, \ln \frac{\tau}{\tau_0}, \\
  & P_T = P_T^{(0)} + P_T^{(1)} = \frac{A^{(0)} + \as^2 \, A^{(1)} - \as^2 \, B^{(1)}}{\tau} - \frac{\as^2 \, B^{(1)}}{\tau} \, \ln \frac{\tau}{\tau_0}, \\
  & P_L = P_L^{(1)} = \frac{\as^2 \, B^{(1)}}{\tau} .
\end{align}
\end{subequations}
Putting the strong coupling to zero in Eqs.~\eqref{eq:results},
$\as=0$, we recover the leading-order classical free-streaming result
\eqref{eq:free} (with some coefficient $A^{(0)}$). The corrections due
to the single iteration of the Boltzmann collision term enter at the
order $\as^2$ in Eqs.~\eqref{eq:results} with some coefficients
$A^{(1)}$ and $B^{(1)}$. (Note that $\tau_0$ is the lower cutoff for
the applicability times of kinetic theory: it is applicable for $\tau
> \tau_0$.) Since the pressure components in kinetic theory must be
positive, we see that $B^{(1)} >0$. Therefore, the
$B^{(1)}$-corrections in \eqref{eq:results} appear to lead to $P_L >0$
and simultaneously make the energy density $\epsilon$ decrease faster
than $1/\tau$. We see that kinetic theory predicts that a single $2
\to 2$ rescattering of two classical fields would generate the first
isotropization correction we are after. The remaining question is
whether such correction arises in the full field-theoretical
calculation.

The perturbative solution of Boltzmann equation from
\cite{WuKovchegov} is quite general and is valid for the $2\to 2$
collision term calculated in any theory. While the discussion in
\cite{WuKovchegov} concentrated on QCD with the expansion in
Eqs.~\eqref{eq:results} being in powers of $\as$, it can be easily
modified to apply to the scalar $\varphi^4$ theory in question here by
simply replacing $\as \to \lambda$ everywhere. (All the appropriate
constants and scattering amplitudes would be modified as well, but
this would not matter so much for us since we are interested mainly in
the form of the $\tau$-dependence of the energy-momentum tensor.)
Equations~\eqref{eq:results} can be rewritten as
\begin{subequations}\label{eq:results_scalar}
\begin{align}
  & \epsilon = \epsilon^{(0)} + \epsilon^{(1)} = \frac{A^{(0)} + \lambda^2 \, A^{(1)}}{\tau} - \frac{\lambda^2 \, B^{(1)}}{\tau} \, \ln \frac{\tau}{\tau_0}, \\
  & P_T = P_T^{(0)} + P_T^{(1)} = \frac{A^{(0)} + \lambda^2 \, A^{(1)} - \lambda^2 \, B^{(1)}}{\tau} - \frac{\lambda^2 \, B^{(1)}}{\tau} \, \ln \frac{\tau}{\tau_0}, \\
  & P_L = P_L^{(1)} = \frac{\lambda^2 \, B^{(1)}}{\tau} .
\end{align}
\end{subequations}
(It is understood that $A^{(0)}$, $A^{(1)}$, $B^{(1)}$ and possibly
even $\tau_0$ are different in Eqs.~\eqref{eq:results} and
\eqref{eq:results_scalar}.) Once again we observe that the kinetic
theory has a specific prediction for the outcome of the single $2 \to
2$ rescattering of two classical fields.\footnote{In this work we only
  consider the solution of the Boltzmann equation in terms of the
  coupling expansion in order to compare with the perturbative
  calculation in quantum field theory. The exact numerical solution of
  the Boltzmann equation in the $\phi^4$ theory can be found in
  Ref.~\cite{Epelbaum:2015vxa}, which favors the isotropization at
  late times.} In the remainder of this paper we will verify the
results \eqref{eq:results_scalar} by an explicit diagrammatic
calculation.


\section{Two-point correlation function with a single rescattering:
  Full diagrammatic calculation in momentum space}
\label{sec:x1x2}

\begin{figure}[htb]
\begin{center}
\includegraphics[width=\textwidth]{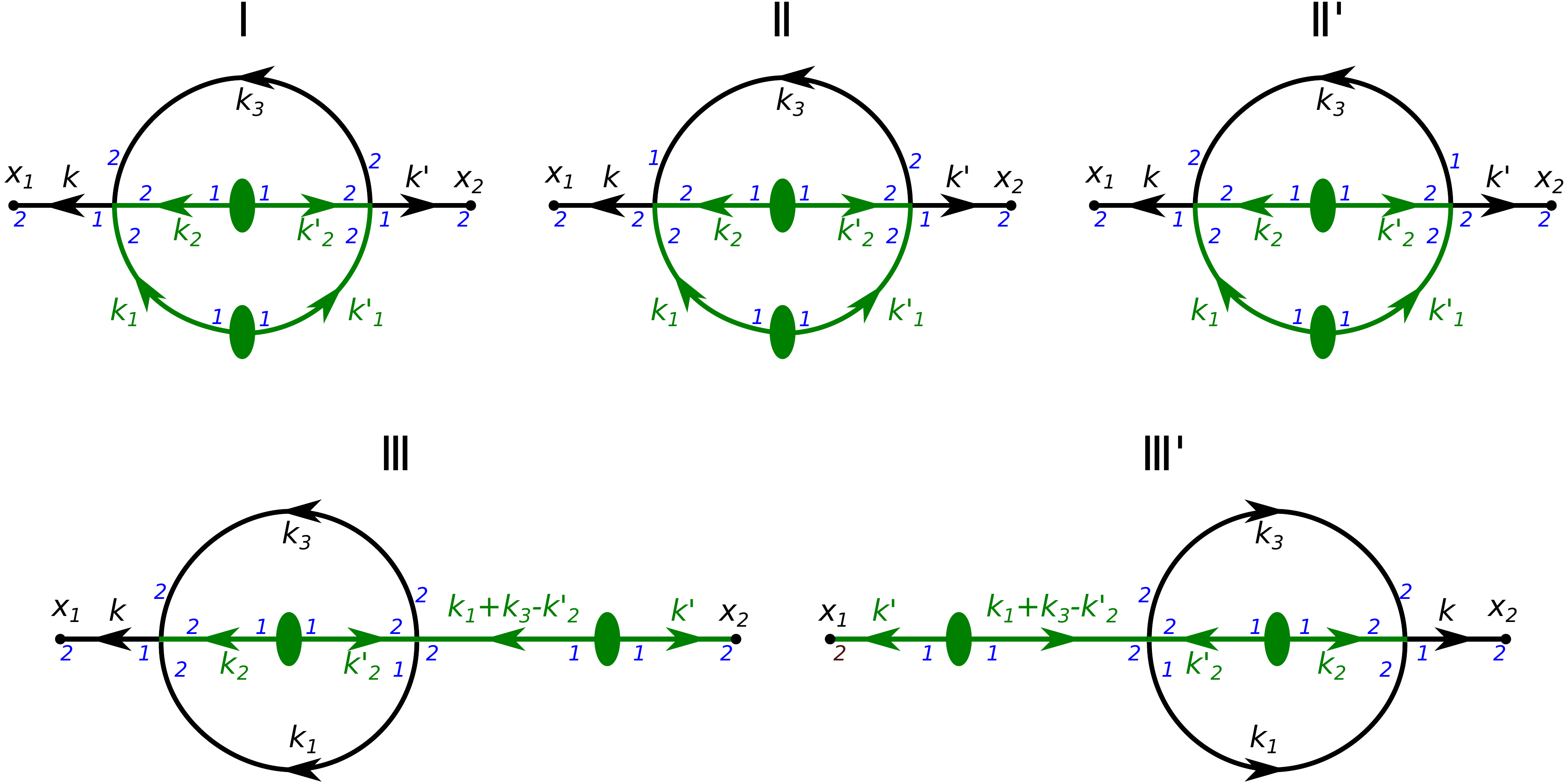}
\end{center}
\caption{The diagrams for the $2 \to 2$ rescattering correction to
  the classical field Green function that we calculate in this
  section. The green oval, along with the attached propagators, denote
  the classical correlation function.}
\label{fig:phi4}
\end{figure}

In this Section our aim is to calculate the late-times $\tau_1,
\tau_2$ asymptotics of the correlation function $G_{22} (x_1, x_2)$
due to a single $2 \to 2$ rescattering involving two classical
fields. The diagrams we want to calculate are shown in
\fig{fig:phi4}. The diagrams are labeled I, II, II', III, and III'
with II' and III' obtainable from II and III by replacing $x_1
\leftrightarrow x_2$ in them. In the kinetic theory language, diagrams
I, II, and II' correspond to the gain term in the (collision term of
the) Boltzmann equation, while diagrams III and III' correspond to the
loss term. Below we will first calculate diagrams I, II, and II'
together, and then calculate diagrams III and III'.

\subsection{Diagrams I, II, and II'}

Let us start with diagram I. In momentum space one has
\begin{align}
  \label{eq:DiagIk1}
  \text{Diag}_\text{I}(k,k') = & - \frac{\lambda^2}{2}\frac{i}{k^2 + i
    \epsilon k^0} \, \frac{i}{k^{\, \prime \, 2} + i \epsilon k^{\,
      \prime \, 0}} \notag \\ & \times \frac{d^4 k_1}{(2 \pi)^4} \,
  \frac{d^4 k'_1}{(2 \pi)^4} \, \frac{d^4 k_2}{(2 \pi)^4} \, \frac{d^4
    k'_2}{(2 \pi)^4} \, \frac{d^4 k_3}{(2 \pi)^4} \, (2 \pi)^4 \,
  \delta^4 (k - k_1 - k_2 -
  k_3) \notag \\
  & \times (2 \pi)^4 \, \delta^4 (k' - k'_1 - k'_2 + k_3) \,
  G^{LO}_{22} (k_1, k'_1) \, G^{LO}_{22} (k_2, k'_2) \, \pi \, \delta
  (k_3^2)
\end{align}
with the leading-order classical correlation function given by
(\ref{eq:Gcl_sc}). Substituting \eqref{eq:Gcl_sc} into
\eqref{eq:DiagIk1} and integrating out all the delta-functions (except
for one) yields
\begin{align}
  \label{eq:DiagIk2}
  \mbox{Diag}_\text{I}(k,k') = & - \frac{\lambda^2}{2} (2\pi)^2 \delta^2(\underline{k}+\underline{k}^{\prime})G_R(k)G_R(k^\prime)\notag\\
  &\times\int\frac{d^2 k_1}{(2 \pi)^2} \, f(k_{1T}) \, f(k_{2T})
  \,\int\frac{d^4 k_3}{(2 \pi)^4} \, \pi
  \, \delta (k_3^2)\notag\\
  &\times I_1 (k^+ - k_3^+, k^- - k_3^-, k_{1T}, k_{2T}) \, I_1 (k'^+
  + k_3^+, k'^- + k_3^-, k_{1T}, k_{2T})
\end{align}
with $\un{k}_2 = \un{k} - \un{k}_1 - \un{k}_3$ and the retarded scalar
Green function
\begin{align}
  \label{eq:Green_ret}
  G_R(p) = \frac{i}{2 (p^++i\epsilon) (p^-+i\epsilon) -
    \underline{p}^2}.
\end{align}

In coordinate space one has
\begin{align}
  \label{eq:DiagI2}
  \mbox{Diagram I} =&\, \int \frac{d^4 k}{(2
    \pi)^4} \,\frac{d^4 k'}{(2 \pi)^4} \, e^{- i k \cdot x_1- i k' \cdot x_2} \,
  \text{Diag}_\text{I}(k,k')\notag\\
  =& - \frac{\lambda^2}{2} \,
  \int\limits_{-\infty}^\infty \frac{d k^+ \, d k^-}{(2 \pi)^2} \,
  e^{- i k^+ x_1^- - i k^- x_1^+} \, \frac{i}{k^2 + i \epsilon k^0} \,
  \int\limits_{-\infty}^\infty \frac{d k'^+ d k'^-}{(2 \pi)^2} \, e^{-
    i k'^+ x^-_2 - i k'^- x_2^+} \notag \\ & \times \frac{i}{k^{\,
      \prime \, 2} + i \epsilon k^{\, \prime \, 0}} \, \frac{d^2 k}{(2
    \pi)^2} \frac{d^2 k_1}{(2 \pi)^2} \, f(k_{1T}) \, f(k_{2T}) \,
  e^{i \un{k} \cdot \un{x}_{12}} \, \frac{d^4 k_3}{(2 \pi)^4} \, \pi
  \, \delta (k_3^2)  \notag \\
  & \times I_1 (k^+ - k_3^+, k^- - k_3^-, k_{1T}, k_{2T}) \, I_1 (k'^+
  + k_3^+, k'^- + k_3^-, k_{1T}, k_{2T}).
\end{align}
From now on it is implicitly understood that
\begin{align}
  \label{eq:kTs}
  \un{k}' = - \un{k}, \ \ \ \un{k}_2 = \un{k} - \un{k}_1 - \un{k}_3.
\end{align}
In arriving at \eq{eq:DiagI2} we have defined
\begin{align}
  \label{I1def}
  I_1(q^+,q^-,p_T, |\un{q} - \un{p}|) = \int\limits_{-\infty}^\infty
  \frac{dp^+ dp^-}{(2\pi)^2} \, G_R(p) \, G_R(q-p).
\end{align}

Equation \eqref{eq:DiagI2} can be written more compactly by defining
\begin{align}
  \label{eq:I3def}
  I_3 (x^+, x^-, k_T, k_{1T}, k_{2T}) = \int\limits_{-\infty}^\infty
  \frac{dk^+dk^-}{(2\pi)^2} \ & e^{-i k^+ x^--ik^-x^+} \, G_R(k)
  \notag \\ & \times I_1(k^+ - k_3^+,k^- - k_3^-, k_{1T},k_{2T}).
\end{align}
We obtain
\begin{align}
  \label{eq:DiagI3}
  \mbox{Diagram I} = & - \frac{\lambda^2}{2} \, \int \frac{d^2 k}{(2
    \pi)^2} \frac{d^2 k_1}{(2 \pi)^2} \, f(k_{1T}) \, f(k_{2T}) \,
  e^{i \un{k} \cdot \un{x}_{12}} \, \frac{d^4 k_3}{(2 \pi)^4} \, \pi
  \, \delta (k_3^2)  \notag \\
  & \times I_3 (x_1^+, x_1^-, k_T, k_{1T}, k_{2T}) \, I_3 (x_2^+,
  x_2^-, k_T, k_{1T}, k_{2T}).
\end{align}

While the exact evaluation of $I_3$ appears to be rather involved, we
can obtain its late-time asymptotics. To do this we start with the
full expression for $I_3$,
\begin{align}
  \label{eq:I3eval1}
  I_3 (x^+, x^-, k_T, k_{1T}, k_{2T}) = & \int\limits_{-\infty}^\infty
  \frac{dk^+dk^-}{(2\pi)^2} \ e^{-i k^+ x^--ik^-x^+} \, G_R(k) \notag
  \\ & \times \int\limits_{-\infty}^\infty \frac{dk_1^+
    dk_1^-}{(2\pi)^2} \, G_R(k_1) \, G_R(k - k_3 - k_1),
\end{align}
and integrate over $k^-$ 
to obtain
\begin{align}
  \label{eq:I3eval2}
  & I_3 (x^+, x^-, k_T, k_{1T}, k_{2T}) = \int\limits_{-\infty}^\infty
  \frac{dk^+}{2\pi} \, \frac{e^{-i k^+ x^- - i \, \frac{k_T^2}{2 (k^+
        + i \epsilon)} x^+}}{2 (k^+ + i \epsilon)} \,
  \int\limits_{-\infty}^\infty \frac{dk_1^+ dk_1^-}{(2\pi)^2} \, G_R
  (k_1) \\ & \times \frac{-i}{2 (k^+-k_3^+ - k_1^+ + i \epsilon)} \,
  \frac{1 - e^{-i \left( k_3^- + k_1^- + \frac{k_{2T}^2}{2 (k^+ -k_3^+
          - k_1^+ + i \epsilon)} - \frac{k_T^2}{2 (k^+ + i
          \epsilon)}\right) \, x^+}}{k_3^- + k_1^- + \frac{k_{2T}^2}{2
      (k^+ -k_3^+ - k_1^+ + i \epsilon)} - \frac{k_T^2}{2 (k^+ + i
      \epsilon)}}, \notag
\end{align}
in a form reminiscent of the light-cone perturbation theory (LCPT)
\cite{Lepage:1980fj}.

Now let us apply the large-$\tau$ limit to \eq{eq:I3eval2}. Note that
since the limit of a product is equal to the product of the limits, we
can first take the large-$x^+$ limit of the last fraction in
\eq{eq:I3eval2}. To do so we observe that
\begin{align}
  \label{eq:identity}
  \lim_{x \to + \infty} \frac{1- e^{-i \, a \, x}}{a} = \frac{1}{a - i
    \, \epsilon}
\end{align}
in the distribution sense. Indeed this is true for any function $h(a)$
of real variable $a$ decomposable into a Fourier integral,
\begin{align}
  \label{eq:h}
  h(a) = \int\limits_{-\infty}^\infty d\xi \, e^{i \, a \, \xi} \,
  {\tilde h} (\xi). 
\end{align}
To see this we simply point out that
\begin{align}
  \label{eq:proof}
  \lim_{x \to + \infty} \int\limits_{-\infty}^\infty da \, e^{i \, a
    \, \xi} \, \frac{1- e^{-i \, a \, x}}{a} & = \lim_{x \to + \infty}
  \int\limits_{-\infty}^\infty da \, e^{i \, a \, \xi} \, \frac{1-
    e^{-i \, a \, x}}{a - i \epsilon} \notag \\ & = 2 \pi i \, \lim_{x
    \to + \infty} [\theta (\xi) - \theta (\xi - x)] = 2 \pi i \,
  \theta (\xi),
\end{align}
which is identical to the same convolution of $e^{i \, a \, \xi}$ with
the right-hand side of \eq{eq:identity},
\begin{align}
  \label{eq:rhs}
  \int\limits_{-\infty}^\infty da \, e^{i \, a \, \xi} \, \frac{1}{a -
    i \, \epsilon} = 2 \pi i \, \theta (\xi).
\end{align}

Applying \eq{eq:identity} to the last fraction in \eq{eq:I3eval2}
after neglecting all the $i \epsilon$'s in the exponent we obtain
\begin{align}
  \label{eq:I3eval3}
  I_3 (x^+, x^-, k_T, & \, k_{1T}, k_{2T}) \bigg|_{\tau \to \infty} =
  \int\limits_{-\infty}^\infty \frac{dk^+}{2\pi} \, \frac{e^{-i k^+
      x^- - i \, \frac{k_T^2}{2 (k^+ + i \epsilon)} x^+}}{2 (k^+ + i
    \epsilon)} \, \int\limits_{-\infty}^\infty \frac{dk_1^+
    dk_1^-}{(2\pi)^2} \, G_R (k_1) \notag \\ & \times \frac{-i}{2
    (k^+-k_3^+ - k_1^+ + i \epsilon)} \, \frac{1}{k_3^- + k_1^- +
    \frac{k_{2T}^2}{2 (k^+ -k_3^+ - k_1^+ + i \epsilon)} - k^- - i
    \epsilon},
\end{align}
where by $k^-$ we now imply its on-shell value,
\begin{align}
  \label{eq:k-}
  k^- = \frac{k_T^2}{2 k^+}. 
\end{align}
Equation \eqref{eq:I3eval3} easily simplifies into
\begin{align}
  \label{eq:I3eval4}
  I_3 (x^+, x^-, k_T, k_{1T}, k_{2T}) \bigg|_{\tau \to \infty} =
  \int\limits_{-\infty}^\infty \frac{dk^+}{2\pi} & \, \frac{e^{-i k^+
      x^- - i \, \frac{k_T^2}{2 (k^+ + i \epsilon)} x^+}}{2 (k^+ + i
    \epsilon)} \notag \\ & \times \, \int\limits_{-\infty}^\infty
  \frac{dk_1^+ dk_1^-}{(2\pi)^2} \, G_R (k_1) \, G_R (k-k_3 - k_1),
\end{align}
which, with the help of \eq{I1def} becomes
\begin{align}
  \label{eq:I3eval5}
  I_3 (x^+, x^-, k_T, k_{1T}, k_{2T}) \bigg|_{\tau \to \infty} =
  \int\limits_{-\infty}^\infty \frac{dk^+}{2\pi} \, \frac{e^{-i k^+
      x^- - i \, \frac{k_T^2}{2 (k^+ + i \epsilon)} x^+}}{2 (k^+ + i
    \epsilon)} \, I_1 (k^+ - k_3^+, k^- - k_3^-, k_{1T}, k_{2T}).
\end{align}
Here again $k^-$ is given by \eq{eq:k-}.

Substituting \eq{eq:I3eval5} into \eq{eq:DiagI3} yields
\begin{align}
  \label{eq:DiagI4}
  \mbox{Diagram I} & \, \bigg|_{\tau_1, \tau_2 \to \infty} = -
  \frac{\lambda^2}{2} \, \int\limits_{-\infty}^\infty
  \frac{dk^+}{2\pi} \, \frac{e^{-i k^+ x_1^- - i \, \frac{k_T^2}{2
        (k^+ + i \epsilon)} x_1^+}}{2 (k^+ + i \epsilon)} \,
  \int\limits_{-\infty}^\infty \frac{dk'^+}{2\pi} \, \frac{e^{-i k'^+
      x_2^- - i \, \frac{k_T^2}{2 (k'^+ + i \epsilon)} x_2^+}}{2 (k'^+
    + i \epsilon)} \notag \\
  & \times \, \frac{d^2 k}{(2 \pi)^2} \frac{d^2 k_1}{(2 \pi)^2} \,
  f(k_{1T}) \, f(k_{2T}) \, e^{i \un{k} \cdot \un{x}_{12}} \,
  \frac{d^4 k_3}{(2 \pi)^4} \, \pi \, \delta
  (k_3^2) \notag \\
  & \times \, I_1 (k^+ - k_3^+, k^- - k_3^-, k_{1T}, k_{2T}) \, I_1
  (k'^+ + k_3^+, k'^- + k_3^-, k_{1T}, k_{2T}).
\end{align}
Note that here and throughout the paper, when writing $\tau \to
\infty$, we mean large but finite proper time $\tau$, that is $\tau \,
p_T \gg 1$ with $p_T$ being any of the transverse momenta in the
problem.

For now we leave the diagram I as evaluated in \eq{eq:DiagI4} and turn
our attention to diagrams II and II'. Employing \eq{eq:Gcl_sc} and
integrating over all delta-functions except for one in diagrams II and
II' gives in momentum space
\begin{align}
  \label{eq:DiagIIk}
  & \mbox{Diag}_{\mbox{II}}(k,k') = - \frac{\lambda^2}{2}(2\pi)^2 \delta^2(\underline{k}+\underline{k}^{\prime})\int\frac{d^2 k_1}{(2 \pi)^2} \, f(k_{1T}) \, f(k_{2T}) \notag\\
  & \times\int\, \frac{d^4 k_3}{(2 \pi)^4} \, \pi\left[\delta (k^2)
    G_A(k_3) G_R(k')+G_R(k)G_R(k_3)\delta(k^{\prime 2})\right]\,
  \notag \\ & \times \, I_1 (k^+ - k_3^+, k^- - k_3^-, k_{1T}, k_{2T})
  \, I_1 (k'^+ + k_3^+, k'^- + k_3^-, k_{1T}, k_{2T}).
\end{align}

In coordinate space we have
\begin{align}
  \label{eq:DiagII2}
  & \mbox{Diagrams II+II'}=\, \int \frac{d^4 k}{(2 \pi)^4} \, \frac{d^4 k'}{(2 \pi)^4} \,e^{- i k \cdot x_1- i k' \cdot x_2} \mbox{Diag}_{\mbox{II}}(k,k') \\
  &= - \frac{\lambda^2}{2} \, \int\limits_{-\infty}^\infty \frac{d
    k^+}{8 \pi |k^+|} \, e^{-i k^+ x_1^- - i \, \frac{k_T^2}{2 k^+}
    x_1^+} \frac{d^2 k}{(2 \pi)^2} \frac{d^2 k_1}{(2 \pi)^2} \,
  f(k_{1T}) \, f(k_{2T}) \, e^{i \un{k} \cdot \un{x}_{12}} \notag \\ &
  \times \, \frac{d^4 k_3}{(2 \pi)^4} \, \frac{i}{k_3^{2} - i \epsilon
    k_3^{0}} \, I_3 (x_2^+, x_2^-, k_T, k_{1T}, k_{2T}) \, I_1 (k^+ -
  k_3^+, k^- - k_3^-, k_{1T}, k_{2T}) + (x_1 \leftrightarrow x_2)
  \notag
\end{align}
with $k^-$ given by \eq{eq:k-}. To study the late-time asymptotics of
diagrams II and II' we employ \eq{eq:I3eval5}. This gives
\begin{align}
  \label{eq:DiagII3}
  & \mbox{Diagrams II+II'} = - \frac{\lambda^2}{2} \,
  \int\limits_{-\infty}^\infty \frac{d k^+}{8 \pi |k^+|} \, e^{-i k^+
    x_1^- - i \, \frac{k_T^2}{2 k^+} x_1^+}
  \int\limits_{-\infty}^\infty \frac{dk'^+}{2\pi} \, \frac{e^{-i k'^+
      x_2^- - i \, \frac{k_T^2}{2 (k'^+ + i \epsilon)} x_2^+}}{2 (k'^+
    + i \epsilon)} \notag \\ & \times \, \frac{d^2 k}{(2 \pi)^2}
  \frac{d^2 k_1}{(2 \pi)^2} \, f(k_{1T}) \, f(k_{2T}) \, e^{i \un{k}
    \cdot \un{x}_{12}} \, \frac{d^4 k_3}{(2 \pi)^4} \,
  \frac{i}{k_3^{2} - i \epsilon k_3^{0}} \notag \\ & \times \, I_1
  (k^+ - k_3^+, k^- - k_3^-, k_{1T}, k_{2T}) \, I_1 (k'^+ + k_3^+,
  k'^- + k_3^-, k_{1T}, k_{2T}) + (x_1 \leftrightarrow x_2).
\end{align}

Adding diagrams I, II, and II' given by Eqs.~\eqref{eq:DiagI4} and
\eq{eq:DiagII3} together we arrive at
\begin{align}
  \label{eq:Diag_all1}
  & \mbox{I+II+II'} \, \bigg|_{\tau_1, \tau_2 \to \infty} = -
  \frac{\lambda^2}{2} \, \int\limits_{-\infty}^\infty
  \frac{dk^+}{2\pi} \, \frac{e^{-i k^+ x_1^- - i \, \frac{k_T^2}{2
        (k^+ + i \epsilon)} x_1^+}}{2 (k^+ + i \epsilon)} \,
  \int\limits_{-\infty}^\infty \frac{dk'^+}{2\pi} \, \frac{e^{-i k'^+
      x_2^- - i \, \frac{k_T^2}{2 (k'^+ + i \epsilon)} x_2^+}}{2 (k'^+
    + i \epsilon)} \notag \\
  & \times \, \frac{d^2 k}{(2 \pi)^2} \frac{d^2 k_1}{(2 \pi)^2} \,
  f(k_{1T}) \, f(k_{2T}) \, e^{i \un{k} \cdot \un{x}_{12}} \,
  \frac{d^4 k_3}{(2 \pi)^4} \notag \\ & \times \, \frac{1}{2} \,
  \left[ 2 \pi \, \delta (k_3^2) + \mbox{Sign} (k^+) \frac{i}{k_3^{2}
      - i \epsilon k_3^{0}} + \mbox{Sign} (k'^+) \frac{i}{k_3^{2} + i
      \epsilon k_3^{0}} \right]
  \notag \\
  & \times \, I_1 (k^+ - k_3^+, k^- - k_3^-, k_{1T}, k_{2T}) \, I_1
  (k'^+ + k_3^+, k'^- + k_3^-, k_{1T}, k_{2T}).
\end{align}

The late-time asymptotics of diagrams I, II and II' given by
\eq{eq:Diag_all1} is dominated by the saddle points of the $k^+$ and
$k'^+$ integrals, unless the integrands have other singularities which
may prevent deforming the integration contours into the steepest
descent contours. Regardless of the analytic structure of the
integrands, we can argue that late-time asymptotics at $x^+_1 = x^+_2,
\, x^-_1 = x^-_2$ (as is needed for calculation of expectation values
of local operators, e.g. of the energy-momentum tensor) is dominated
by the regions of integration where $k^+ = - k'^+$: indeed, the
dominant values of $k^+$ and $k'^+$ have to be such that the two
oscillating exponentials in \eq{eq:Diag_all1},
\begin{align}
  \label{eq:exps}
  e^{-i k^+ x_1^- - i \, \frac{k_T^2}{2 (k^+ + i \epsilon)} x_1^+} \,
  e^{-i k'^+ x_2^- - i \, \frac{k_T^2}{2 (k'^+ + i \epsilon)} x_2^+},
\end{align}
cancel each other (for $x^+_1 = x^+_2, \, x^-_1 = x^-_2$), giving no
oscillations in the end. (In fact, functions that oscillate rapidly at
late times can be simply neglected in determining the asymptotics.)
Therefore, expecting $k^+ = - k'^+$, we can put
\begin{align}
  \label{eq:signs}
  \mbox{Sign} (k^+) = - \mbox{Sign} (k'^+)
\end{align}
in \eq{eq:Diag_all1}. This leads to 
\begin{align}
  \label{eq:simpl}
  & \left[ 2 \pi \, \delta (k_3^2) + \mbox{Sign} (k^+)
    \frac{i}{k_3^{2} - i \epsilon k_3^{0}} + \mbox{Sign} (k'^+)
    \frac{i}{k_3^{2} + i \epsilon k_3^{0}} \right] \Bigg|_{\mbox{Sign}
    (k^+) = - \mbox{Sign} (k'^+)} \notag \\ & = 4 \pi \, \delta
  (k_3^2) \, \theta (- k^+ \, k'^+) \, \theta (- k^+ \, k_3^+)
\end{align}
and the sum of the diagrams I, II, and II' at late times becomes
\begin{align}
  \label{eq:Diag_all2}
  & \mbox{I+II+II'} \, \bigg|_{\tau_1, \tau_2 \to \infty} = -
  \frac{\lambda^2}{2} \, \int\limits_{-\infty}^\infty
  \frac{dk^+}{2\pi} \, \frac{e^{-i k^+ x_1^- - i \, \frac{k_T^2}{2
        (k^+ + i \epsilon)} x_1^+}}{2 (k^+ + i \epsilon)} \,
  \int\limits_{-\infty}^\infty \frac{dk'^+}{2\pi} \, \frac{e^{-i k'^+
      x_2^- - i \, \frac{k_T^2}{2 (k'^+ + i \epsilon)} x_2^+}}{2 (k'^+
    + i \epsilon)} \notag \\
  & \times \, \frac{d^2 k}{(2 \pi)^2} \frac{d^2 k_1}{(2 \pi)^2} \,
  f(k_{1T}) \, f(k_{2T}) \, e^{i \un{k} \cdot \un{x}_{12}} \,
  \frac{d^4 k_3}{(2 \pi)^4} \, 2 \pi \, \delta (k_3^2) \, \theta (-
  k^+ \, k'^+) \, \theta (- k^+ \, k_3^+)
  \notag \\
  & \times \, I_1 (k^+ - k_3^+, k^- - k_3^-, k_{1T}, k_{2T}) \, I_1
  (k'^+ + k_3^+, k'^- + k_3^-, k_{1T}, k_{2T}).
\end{align}
Such partial cancellation between the diagrams I, II, and II' was seen
before in the framework of kinetic theory\cite{Mueller:2002gd}.

Further evaluation of the expression \eqref{eq:Diag_all2} appears to
be impossible without an explicit expression for $I_1$. The
corresponding calculation is carried out in Appendix~\ref{sec:I1}
resulting in
\begin{align}
  \label{eq:I1final}
  I_1 (q^+, q^-, k_T, p_T ) = \frac{1}{4 q^-\, \left[q^+ -
      \frac{(k_T+p_T)^2}{2q^- } \right]^\frac{1}{2} \, \left[q^+ -
      \frac{(k_T-p_T)^2}{2q^- } \right]^\frac{1}{2}},
\end{align}
where the branch cut of the square root is chosen along the negative
imaginary axis for the later convenience. Note also that the sum
$k_T+p_T$ and the difference $k_T-p_T$ involve the magnitudes of the
transverse momenta, and are not a sum or a difference of vectors.

\begin{figure}[htb]
\begin{center}
\includegraphics[width=\textwidth]{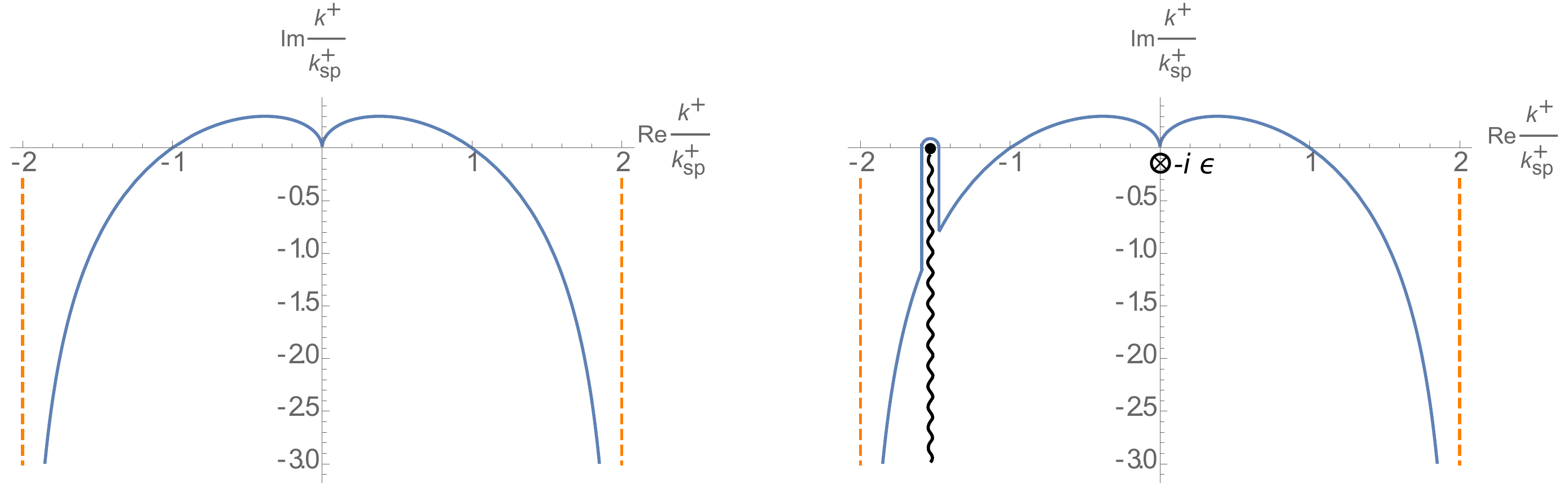}
\end{center}
\caption{Left panel shows the steepest descent contour for the $k^+$
  integral.  Vertical dashed lines denote the asymptotics of the
  steepest descent contour. The right panel shows a sample contour
  which may result from trying to distort the $k^+$ integration
  contour from going along the real axis to the steepest descent
  contour (see text).}
\label{fig:steep}
\end{figure}

Once again let us point out that to obtain the late-time asymptotics
of \eq{eq:Diag_all2} we need to try to deform the contours of the
$k^+$ and $k'^+$ integrals into the steepest descent shape, in order
to perform the saddle point approximation. This contour deformation
may be affected by the presence of singularities in the $k^+$ and
$k'^+$ complex planes. For definitiveness, consider the $k^+$
integral. For all other momenta in \eq{eq:Diag_all2} fixed, it has an
essential singularity at $k^+ = - i \epsilon$ and branch cuts due to
$I_1 (k^+ - k_3^+, k^- - k_3^-, k_{1T}, k_{2T})$. The steepest descent
contour for the $k^+$ integral is shown in the left panel of
\fig{fig:steep}. The saddle points of the $k^+$ integral are given by
$k^+ = \pm k^+_{sp}$ with
\begin{align}
  \label{eq:ksp}
  k^+_{sp} = \frac{k_T}{\sqrt{2}} \, \sqrt{\frac{x^+_1}{x^-_1}}.
\end{align}
They correspond to points $(\pm 1, 0)$ in both panels of
\fig{fig:steep}. In the case of no singularities in the complex $k^+$,
it is clear that one can easily deform the $k^+$ integration from
running along the real axis to the steepest descent curve in the left
panel of \fig{fig:steep}.

The right panel of \fig{fig:steep} illustrates what happens if one
tries to deform the real axis contour into the steepest descent one in
the presence of singularities in the complex $k^+$ plane. In that plot
we explicitly show the essential singularity at $k^+ = - i \epsilon$
in the integrand of \eq{eq:Diag_all2}: one can see that it does not
interfere with the integration contour deformation into the steepest
descent shape because it lies outside the contour, and, when the
contour approaches this singularity, it does so along the positive
imaginary axis near the origin, $k^+ \approx + i \epsilon'$. This
region of integration is exponentially suppressed as one can see by
plugging $k^+ \approx + i \epsilon'$ into the first exponential of
\eq{eq:Diag_all2}. Hence we do not need to worry about the singularity
at $k^+ = - i \epsilon$.

In contrast, the branch cuts may possibly interfere with the contour
deformation: this is illustrated in the right panel of \fig{fig:steep}
by a sample vertical branch cut with the branch point on the real
axis. This branch cut is not an accurate representation of the branch
cuts of the $k^+$ integrand, and is shown here as a toy model to
illustrate the possibility that in deforming the integration contour
one may have to wrap the contour around a part of the branch cut as
well. The corresponding sample integral would look like
(cf. Eqs.~\eqref{eq:Diag_all2} and \eqref{eq:I1final})
\begin{align}
  \label{eq:sample1}
  \int\limits_{-\infty}^\infty \frac{dk^+}{2\pi} \, \frac{e^{-i k^+
      x_1^- - i \, \frac{k_T^2}{2 (k^+ + i \epsilon)} x_1^+}}{2 (k^+ +
    i \epsilon) \, \sqrt{k^+ - k^+_{br} + i \epsilon}}
\end{align}
with $k^+_{br} - i \epsilon$ the branch point near the real axis, as
shown in the right panel of \fig{fig:steep}. (Note again the the
branch cut of the square root is chosen along the negative imaginary
axis.) Writing
\begin{align}
  \label{eq:brcont}
  k^+ = k^+_{br} - i y
\end{align}
with some real variable $y$ we can approximate the contribution to
the integral in \eq{eq:sample1} from the part of the contour wrapped
around the branch cut by
\begin{align}
  \label{eq:sample2}
  \approx - i \, \int\limits_{0}^\infty \frac{dy}{2\pi} \, \frac{e^{-
      y \, \left( x_1^- - \frac{k_T^2}{2 (k_{br}^+)^2} \, x_1^+
      \right) -i k^+_{br} x_1^- - i \, \frac{k_T^2}{2 k^+_{br}}
      x_1^+}}{2 k^+_{br} \, \sqrt{- i y}} = \frac{1-i}{\sqrt{2}} \
  \frac{e^{-i k^+_{br} x_1^- - i \, \frac{k_T^2}{2 k^+_{br}} x_1^+}}{4
    \sqrt{\pi} \, k^+_{br} \, \sqrt{x_1^- - \frac{k_T^2}{2
        (k_{br}^+)^2} \, x_1^+}}.
\end{align}
In the process we have assumed that the branch cut section of the
contour is long enough for the upper limit of the $y$ integral in
\eq{eq:sample2} to be replaced by infinity: this is only valid if
$k^+_{br}$ is sufficiently far from the saddle point $k^+_{sp}$ (and
for $|k^+_{br}| > |k^+_{sp}|$). More specifically, we need $|k^+_{br}
- k^+_{sp}| \gg 1/x_1^- \sim 1/\tau_1$. Since $ 1/\tau_1$ is small for
large times $\tau_1$, this assumption is justified in most cases.

What we learn from the sample integration in Eqs.~\eqref{eq:sample1}
and \eqref{eq:sample2} is that the branch cut contribution is
dominated by the branch point, $k^+ = k^+_{br}$, with a small region
$y \in [0, \sim 1/\tau_1]$ near the branch point contributing
dominantly to (this part of) the integral.

We are now ready to tackle the full $k^+$ and $k'^+$ integrals in
\eq{eq:Diag_all2}. First we need to identify the branch points and
branch cuts of the $k^+$ and $k'^+$ integrals. Starting with the $k^+$
integral we see that its branch cuts originate in $I_1 (k^+ - k_3^+,
k^- - k_3^-, k_{1T}, k_{2T})$ as follows from
\eq{eq:I1final}. Defining
\begin{align}
  \label{eq:ratio}
  \xi = \frac{k^+}{k_3^+}
\end{align}
we conclude that there are four branch points given by 
\begin{align}
  \label{eq:brpts}
  \xi_{1,2,3,4} = \frac{1}{2 k_{3T}^2} \, \left[ k_T^2 + k_{3T}^2 -
    (k_{1T} \pm k_{2T})^2 \pm \sqrt{(k_T^2 + k_{3T}^2 - (k_{1T} \pm
      k_{2T})^2)^2 - 4 k_T^2 \, k_{3T}^2} \right].
\end{align}
(The sign in $(k_{1T} \pm k_{2T})$ is either a plus or a minus
simultaneously inside the square root and outside: one cannot have
$k_{1T} + k_{2T}$ in one place in the expression and $k_{1T} - k_{2T}$
in another.)  These branch points can be real or complex. In the
latter case of complex $\xi$ the branch point may contribute to the
integral only if $|\mbox{Re} \, \xi| > |k^+_{sp}|$, as follows from
the contour in \fig{fig:steep}. In such case one can easily show that
the contribution of the complex-valued branch point in the lower $k^+$
half-plane is exponentially suppressed. Since for positive $x_1^-$
that we are interested in one needs to close the $k^+$ integration
contour in the lower half-plane, we conclude that we can discard the
contributions of the complex-valued branch points.

We are left with the case of real-values branch points $\xi$. Due to
the presence of $\theta (- k^+ \, k_3^+)$ in \eq{eq:Diag_all2}, only
the negative real $\xi$ can contribute (otherwise the $k^+$
integration never approaches the branch point for it to contribute). A
quick analysis of \eq{eq:brpts} shows that only two solutions can be
real and negative. Let us denote them $\xi_1$ and $\xi_2$, such that
\begin{subequations}
\begin{align}
  \label{eq:brpts2}
  \xi_{1} = \frac{1}{2 k_{3T}^2} \, \left[ k_T^2 + k_{3T}^2 - (k_{1T}
    + k_{2T})^2 + \sqrt{(k_T^2 + k_{3T}^2 - (k_{1T} +
      k_{2T})^2)^2 - 4 k_T^2 \, k_{3T}^2} \right] \\
  \xi_{2} = \frac{1}{2 k_{3T}^2} \, \left[ k_T^2 + k_{3T}^2 - (k_{1T}
    + k_{2T})^2 - \sqrt{(k_T^2 + k_{3T}^2 - (k_{1T} + k_{2T})^2)^2 - 4
      k_T^2 \, k_{3T}^2} \right]
\end{align}
\end{subequations}
with $\xi_2 < \xi_1 <0$. The corresponding branch point in the $k^+$
plane are given by $k^+ = \xi_1 \, k_3^+$ and $k^+ = \xi_2 \,
k_3^+$. Note that the branch points of the $k'^+$ integral are given
by the branch points of $I_1 (k'^+ + k_3^+, k'^- + k_3^-, k_{1T},
k_{2T})$ resulting in $k'^+ = - \xi_1 \, k_3^+$ and $k'^+ = - \xi_2 \,
k_3^+$. 

\begin{figure}[htb]
\begin{center}
  \includegraphics[width=0.4 \textwidth]{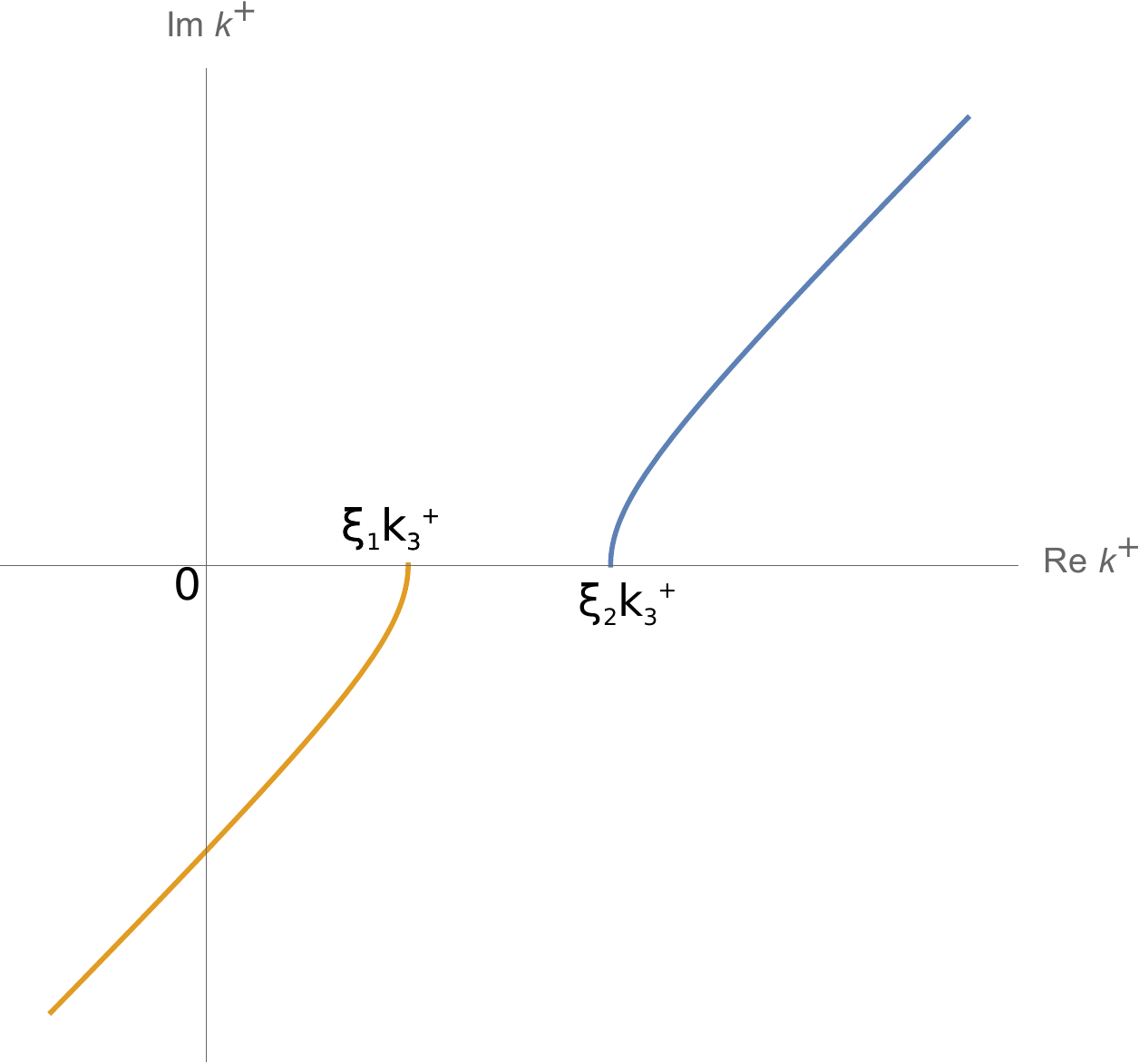}
  \hspace*{3mm} \includegraphics[width=0.4 \textwidth]{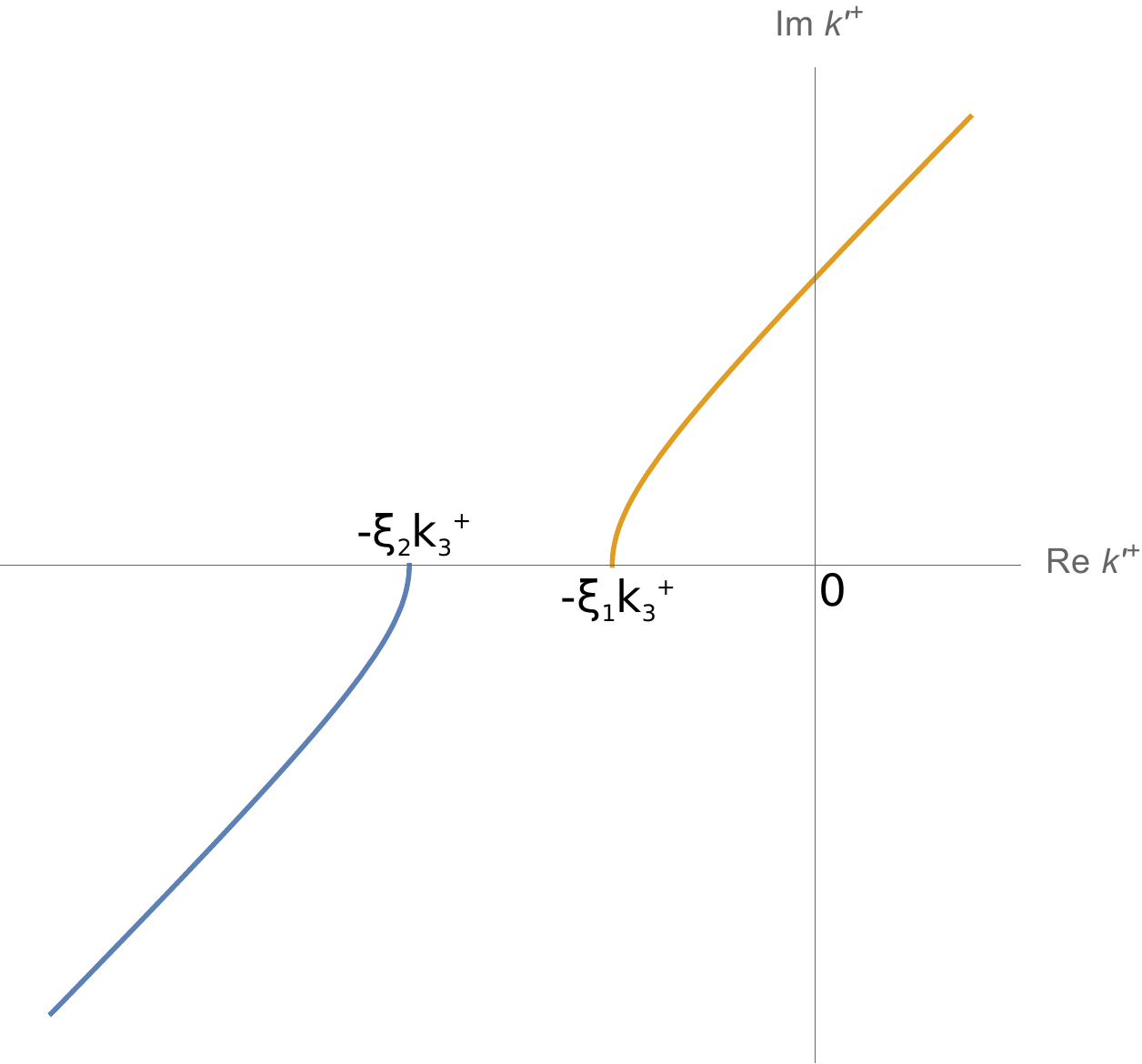}
\end{center}
\caption{The branch cuts of the integrand in \eq{eq:Diag_all2} in the
  complex $k^+$ (left panel) and $k'^+$ (right panel) planes, drawn
  here for the $k_3^+ < 0$ case.}
\label{fig:branch}
\end{figure}

Here we will consider the case of $k_3^+ < 0$ (with $k^+ > 0, k'^+
<0$): the $k_3^+ > 0$ case can be done by analogy. The branch cuts of
the $k^+$ and $k'^+$ integral in the $k_3^+ < 0$ case are shown in
\fig{fig:branch}.

Remembering from the example above that the integrals around our
branch cuts are dominated by the branch points, and invoking the
argument that the late-time asymptotics is dominated by $k'^+ = - k^+$
to avoid rapid oscillations (which would make the function practically
zero), we conclude that if the $k^+$ integral is wrapped around the
branch cut originating at, say, the $\xi_1 k_3^+$ branch point, the
$k'^+$ integral must be wrapped around a branch cut originating at the
$-\xi_1 k_3^+$ branch point. However, as one can see from
\fig{fig:branch}, while the branch cut starting at $\xi_1 k_3^+$ in
the $k^+$ complex plane lies in the lower half-plane, the branch cut
starting at $-\xi_1 k_3^+$ in the $k'^+$ complex plane lies in the
upper half-plane. For $x_1^-, x_2^- > 0$ that we are interested in one
needs to close the $k^+$ and $k'^+$ contours in the lower half-plane:
hence one cannot simultaneously pick up the contributions of the
branch cuts originating at $\xi_1 k_3^+$ and $-\xi_1 k_3^+$ (or at
$\xi_2 k_3^+$ and $-\xi_2 k_3^+$) in the $k^+$ and $k'^+$
integrals.\footnote{While it is possible to choose the transverse
  momenta $\un{k}, \un{k_1}, \un{k}_2, \un{k}_3$ to make $\xi_1 =
  \xi_2$, this appears to be a single regular point in the integrand,
  not enhanced by any sort of singularity.} We thus conclude that both
the $k^+$ and $k'^+$ integrals can not be dominated by branch cuts.

For a given value of $k_3^+$, and for $x_1^\pm \approx x_2^\pm$, the
branch points are either near $|k^+_{br} - k^+_{sp}| \lsim 1/\tau$ or
far $|k^+_{br} - k^+_{sp}| \gg 1/\tau$ from the saddle point. The
`near' region is small at large $\tau$ and its contribution is
suppressed by an extra power of $1/\tau$. Hence the leading
contribution to the integral in \eq{eq:Diag_all2} comes from the `far'
region of $|k^+_{br} - k^+_{sp}| \gg 1/\tau$. In this region branch
cuts are clearly separated from the saddle points: if we do not want
to have rapidly oscillating exponentials, we can either have the
contribution of saddle points in both the $k^+$ and $k'^+$ integrals
or of the branch points in both the $k^+$ and $k'^+$ integrals, but
due to the separation of branch cuts and saddle points we cannot have
a mix where the branch point contributes in one and the saddle point
contributes in the other. 
Since we have just eliminated the contribution of the branch points in
both the $k^+$ and $k'^+$ integrals, we are left with the
contributions of saddle points only. The saddle point $k^+$-integral
gives
\begin{align}
  \label{eq:saddle_eval0}
  & \int\limits_{0}^\infty \frac{dk^+}{2\pi} \, \frac{e^{-i k^+ x_1^-
      - i \, \frac{k_T^2}{2 (k^+ + i \epsilon)} x_1^+}}{2 (k^+ + i
    \epsilon)} \, I_1 (k^+ - k_3^+, k^- - k_3^-, k_{1T}, k_{2T})
  \Bigg|_{\tau_1 \to \infty} \notag \\ & \approx e^{- i k_T \, \tau_1}
  \, \frac{1}{2 \pi} \, \frac{1-i}{2} \, \sqrt{\frac{\pi}{k_T \,
      \tau_1}} \, I_1 \left( \frac{k_T}{\sqrt{2}} \, e^{\eta_1} -
    k_3^+, \frac{k_T}{\sqrt{2}} \, e^{-\eta_1} - k_3^-, k_{1T}, k_{2T}
  \right)
\end{align}
and
\begin{align}
  \label{eq:saddle_eval1}
  & \int\limits_{-\infty}^0 \frac{dk'^+}{2\pi} \, \frac{e^{-i k'^+
      x_2^- - i \, \frac{k_T^2}{2 (k'^+ + i \epsilon)} x_2^+}}{2 (k'^+
    + i \epsilon)} \, I_1 (k'^+ + k_3^+, k'^- + k_3^-, k_{1T}, k_{2T})
  \Bigg|_{\tau_2 \to \infty} \notag \\ & \approx - e^{i k_T \, \tau_2}
  \, \frac{1}{2 \pi} \, \frac{1+i}{2} \, \sqrt{\frac{\pi}{k_T \,
      \tau_2}} \, I_1 \left( -\frac{k_T}{\sqrt{2}} \, e^{\eta_2} +
    k_3^+, - \frac{k_T}{\sqrt{2}} \, e^{-\eta_2} + k_3^-, k_{1T},
    k_{2T} \right).
\end{align}

Substituting Eqs.~\eqref{eq:saddle_eval0} and \eqref{eq:saddle_eval1}
into \eq{eq:Diag_all2} and adding the $k_3^+>0$ contribution
yields\footnote{Note that the $k_3^+$ integral in \eq{eq:Diag_all3} is
  finite, as can be checked explicitly.}
\begin{align}
  \label{eq:Diag_all3}
  & \mbox{I+II+II'} \, \bigg|_{\tau_1, \tau_2 \to \infty} =
  \frac{\lambda^2}{16 \pi \sqrt{\tau_1 \, \tau_2}} \, \int \frac{d^2
    k}{(2 \pi)^2} \frac{d^2 k_1}{(2 \pi)^2} \frac{d^2 k_3}{(2 \pi)^2}
  \, f(k_{1T}) \, f(k_{2T}) \, e^{i \un{k} \cdot \un{x}_{12}} \,
  \frac{1}{k_T}
  \notag \\
  & \times \, \Bigg\{ - \int\limits_{-\infty}^0 \frac{d k_3^+}{4 \pi
    k_3^+}\, e^{- i k_T \, (\tau_1 - \tau_2)} \, I_1 \left(
    \frac{k_T}{\sqrt{2}} \, e^{\eta_1} - k_3^+, \frac{k_T}{\sqrt{2}}
    \, e^{-\eta_1} - \frac{\un{k}_3^2}{2 k_3^+}, k_{1T}, k_{2T} \right) \notag \\
  & \times \, I_1 \left( -\frac{k_T}{\sqrt{2}} \, e^{\eta_2} + k_3^+,
    - \frac{k_T}{\sqrt{2}} \, e^{-\eta_2} + \frac{\un{k}_3^2}{2
      k_3^+}, k_{1T}, k_{2T} \right) + \int\limits^{\infty}_0 \frac{d
    k_3^+}{4 \pi k_3^+}\, e^{i k_T \, (\tau_1 - \tau_2)} \notag \\
  & \times \, I_1 \left( -\frac{k_T}{\sqrt{2}} \, e^{\eta_1} - k_3^+,
    - \frac{k_T}{\sqrt{2}}
    \, e^{-\eta_1} - \frac{\un{k}_3^2}{2 k_3^+}, k_{1T}, k_{2T} \right) \notag \\
  & \times \, I_1 \left( \frac{k_T}{\sqrt{2}} \, e^{\eta_2} + k_3^+,
    \frac{k_T}{\sqrt{2}} \, e^{-\eta_2} + \frac{\un{k}_3^2}{2 k_3^+},
    k_{1T}, k_{2T} \right) \Bigg\}.
\end{align}
We conclude that
\begin{align}
  \label{eq:asymI&II}
  \mbox{I+II+II'} \Bigg|_{\tau_1 = \tau_2 = \tau \to \infty} \sim
  \frac{1}{\tau},
\end{align}
consistent with free streaming.


\subsection{Diagrams III and III'}

Our aim now is to calculate diagrams III and III' from
\fig{fig:phi4}. Following the steps we made for other diagrams, we
write
\begin{align}
  \label{eq:DiagIII1}
  & \mbox{Diagrams III+III'} = - \lambda^2 \, \int \frac{d^4
    k}{(2 \pi)^4} \, e^{- i k \cdot x_1} \, \frac{i}{k^{2} + i
    \epsilon k^{0}} \, \int \frac{d^4 k'}{(2 \pi)^4} \, e^{- i k'
    \cdot x_2} \\ & \times \frac{d^4 k_1}{(2 \pi)^4} \, \frac{d^4
    k_2}{(2 \pi)^4} \, \frac{d^4 k'_2}{(2 \pi)^4} \, \frac{d^4 k_3}{(2
    \pi)^4} \, (2 \pi)^4 \, \delta^4 (k - k_1 - k_2 -
  k_3) \notag \\
  & \times \, G^{LO}_{22} (k_1 + k_3 - k'_2 , k') \, G^{LO}_{22} (k_2,
  k'_2) \, \frac{i}{k_1^{2} + i \epsilon k_1^{0}} \, \pi \, \delta
  (k_3^2) + (x_1 \leftrightarrow x_2). \notag
\end{align}
Substituting the leading-order correlators from \eq{eq:Gcl_sc},
integrating out all but one delta-function, and also integrating over
$k'^+$ and $k'^-$ explicitly yields
\begin{align}
  \label{eq:DiagIII2}
  & \mbox{Diagrams III+III'} = \frac{i \, \lambda^2}{2} \,
  \int\limits_{-\infty}^\infty \frac{d k^+ dk^-}{(2 \pi)^2} \, e^{- i
    k^+ x_1^- - i k^- x_1^+} \, \frac{i}{k^{2} + i \epsilon k^{0}} \,
  J_0 (k_T \, \tau_2) \\ & \times \frac{d^2 k}{(2 \pi)^2} \frac{d^2
    k_1}{(2 \pi)^2} \, f(k_{T}) \, f(k_{2T}) \, e^{i \un{k} \cdot
    \un{x}_{12}} \, \frac{d^4 k_3}{(2 \pi)^4} \, \pi \, \delta (k_3^2)
  \, \frac{d k_1^+ dk_1^-}{(2 \pi)^2} \, I_1 (k_1^+ + k_3^+, k_1^- +
  k_3^-, k_{2T}, k_T) \notag \\ & \times \, \frac{i}{k_1^{2} + i
    \epsilon k_1^{0}} \, \frac{1}{(k-k_1 -k_3)^2 + i \epsilon (k^0 -
    k_1^0 - k_3^0)} + (x_1 \leftrightarrow x_2).  \notag
\end{align}
Integrating over $k^-$ and employing the limit from \eq{eq:identity}
one arrives at
\begin{align}
  \label{eq:DiagIII3}
  & \mbox{Diagrams III+III'} = \frac{i \, \lambda^2}{2} \,
  \int\limits_{-\infty}^\infty \frac{d k^+}{2 \pi} \, \frac{e^{- i k^+
      x_1^- - i \frac{k_T^2}{2 (k^+ + i \epsilon)} x_1^+}}{2 (k^+ + i
    \epsilon)} \, J_0 (k_T \, \tau_2) \\ & \times \frac{d^2 k}{(2
    \pi)^2} \frac{d^2 k_1}{(2 \pi)^2} \, f(k_{T}) \, f(k_{2T}) \, e^{i
    \un{k} \cdot \un{x}_{12}} \, \frac{d^2 k_3}{(2 \pi)^2} \, \frac{d
    q^+ dq^-}{(2 \pi)^2} \, I_1 (q^+, q^-, k_{2T}, k_T) \, I_4 (q^+,
  q^-, k_{3T}, k_{1T}) \notag \\ & \times \frac{i}{2 (q^+ - k^+ - i
    \epsilon) \, \left( q^- - \frac{k_T^2}{2 (k^+ + i \epsilon)} - i
      \epsilon\right) - (\un{k} - \un{k}_1 - \un{k}_3)^2} + (x_1
  \leftrightarrow x_2), \notag
\end{align}
where we have defined $q^\pm \equiv k_1^\pm + k_3^\pm$ along with
\begin{align}
  \label{I4def}
  I_4(q^+,q^-,p_T, |\un{q} - \un{p}|) = \int\limits_{-\infty}^\infty
  \frac{dp^+ dp^-}{(2\pi)^2} \, \pi \delta (p^2) \, G_R(q-p).
\end{align}

Consider the integral over $q^+$ and $q^-$ in \eq{eq:DiagIII3}:
\begin{align}
  \label{eq:integral}
  & J(k^+, k_T, k_{1T}, k_{2T}, k_{3T})= \int\limits_{-\infty}^\infty
  \, \frac{d q^+ dq^-}{(2 \pi)^2} \, I_1 (q^+, q^-, k_{2T}, k_T) \,
  I_4 (q^+, q^-, k_{3T}, k_{1T}) \notag \\ & \times \frac{i}{2 (q^+ -
    k^+ - i \epsilon) \, \left( q^- - \frac{k_T^2}{2 (k^+ + i
        \epsilon)} - i \epsilon\right) - (\un{k} - \un{k}_1 -
    \un{k}_3)^2}.
\end{align}
This object is boost-invariant. It is a function of several transverse
momenta, and of only one longitudinal four-vector component -- of
$k^+$. A boost-invariant object cannot depend on only one $k^+$:
hence, it must be independent of $k^+$,\footnote{In principle $J$ may
  still depend on Sign$(k^+)$. While such dependence would slightly
  modify the integration below along with \eq{eq:DiagIII5}, it will
  not change the fact that the $k^+$-integral is dominated by the
  saddle point, and would still lead to the conclusion
  \eqref{eq:asymIII}.} that is,
\begin{align}
  \label{eq:Jsimpl}
  J(k^+, k_T, k_{1T}, k_{2T}, k_{3T})=J(k_T, k_{1T}, k_{2T}, k_{3T}).
\end{align}
\eq{eq:DiagIII3} becomes
\begin{align}
  \label{eq:DiagIII4}
  & \mbox{Diagrams III+III'} = \frac{i \, \lambda^2}{2} \,
  \int\limits_{-\infty}^\infty \frac{d k^+}{2 \pi} \, \frac{e^{- i k^+
      x_1^- - i \frac{k_T^2}{2 (k^+ + i \epsilon)} x_1^+}}{2 (k^+ + i
    \epsilon)} \, J_0 (k_T \, \tau_2) \\ & \times \frac{d^2 k}{(2
    \pi)^2} \frac{d^2 k_1}{(2 \pi)^2} \, f(k_{T}) \, f(k_{2T}) \, e^{i
    \un{k} \cdot \un{x}_{12}} \, \frac{d^2 k_3}{(2 \pi)^2} \, J(k_T,
  k_{1T}, k_{2T}, k_{3T}) + (x_1 \leftrightarrow x_2). \notag
\end{align}
The exact form of $J(k_T, k_{1T}, k_{2T}, k_{3T})$ is not important
for the late-time asymptotics since now we can integrate over $k^+$
exactly, obtaining
\begin{align}
  \label{eq:DiagIII5}
  & \mbox{Diagrams III+III'} = \frac{\lambda^2}{4} \, \int \frac{d^2
    k}{(2 \pi)^2} \frac{d^2 k_1}{(2 \pi)^2} \, J_0 (k_T \, \tau_1) \,
  J_0 (k_T \, \tau_2) \\ & \times f(k_{T}) \, f(k_{2T}) \, e^{i \un{k}
    \cdot \un{x}_{12}} \, \frac{d^2 k_3}{(2 \pi)^2} \, J(k_T, k_{1T},
  k_{2T}, k_{3T}) + (x_1 \leftrightarrow x_2). \notag
\end{align}

We conclude that
\begin{align}
  \label{eq:asymIII}
  \mbox{Diagrams III+III'} \Bigg|_{\tau_1 = \tau_2 = \tau \to \infty}
  \sim \frac{1}{\tau},
\end{align}
again consistent with free streaming.


\subsection{Energy-Momentum Tensor}

To cross-check our results let us calculate the longitudinal pressure
at $z=0$ (in the massless $\varphi^4$ theory at hand):
\begin{align}
  T^{33} & = \left\langle (\partial_z \varphi)^2 + {\cal L}
  \right\rangle \approx \left\langle \frac{1}{2} (\partial_z
    \varphi)^2 + \frac{1}{2} (\partial_t \varphi)^2 - \frac{1}{2}
    (\un{\nabla} \varphi)^2 \right\rangle \notag \\ & = - \frac{1}{2}
  \int \frac{d^4 k}{(2 \pi)^4} \, \frac{d^4 k'}{(2 \pi)^4} \, e^{-i \,
    (k+k') \cdot x} \, \left[ k^z \, k'^z + k^0 \, k'^0 - \un{k} \cdot
    \un{k}' \right] \, G_{22} (k, k') \notag \\ & = - \frac{1}{2} \int
  \frac{d^4 k}{(2 \pi)^4} \, \frac{d^4 k'}{(2 \pi)^4} \, e^{-i \,
    (k+k') \cdot x} \, \left[ k^+ \, k'^+ + k^- \, k'^- - \un{k} \cdot
    \un{k}' \right] \, G_{22} (k, k').
\end{align}
At the leading saddle point $k^+_{sp} = \pm \frac{k_T}{\sqrt{2}} \,
e^{\eta_1}, \ \ k'^+_{sp} = \mp \frac{k_T}{\sqrt{2}} \, e^{\eta_2}$ and
at mid-rapidity ($z=0$) the square brackets become
\begin{align}
  \left[ k^+ \, k'^+ + k^- \, k'^- - \un{k} \cdot \un{k}' \right]
  \bigg|_{z=0} = \left[ - \frac{k_T^2}{2} \, e^{\eta_1 - \eta_2} -
    \frac{k_T^2}{2} \, e^{\eta_2 - \eta_1} + k_T^2 \right]
  \bigg|_{\eta_1 = \eta_2=0} =0.
\end{align}
(We have also put $\un{k}' = - \un{k}$.) We see that the leading
saddle point contribution gives the longitudinal pressure 
\begin{align}
  \label{eq:pLfin}
  P_L \big|_{\tau \gg \frac{1}{Q_s}} =0,
\end{align}
at late time, as characteristic of free streaming.

For completeness, let us calculate the energy density:
\begin{align}
  T^{00} & = \left\langle (\partial_t \varphi)^2 - {\cal L}
  \right\rangle \approx \left\langle \frac{1}{2} (\partial_z
    \varphi)^2 + \frac{1}{2} (\partial_t \varphi)^2 + \frac{1}{2}
    (\un{\nabla} \varphi)^2 \right\rangle \notag \\ & = - \frac{1}{2}
  \int \frac{d^4 k}{(2 \pi)^4} \, \frac{d^4 k'}{(2 \pi)^4} \, e^{-i \,
    (k+k') \cdot x} \, \left[ k^z \, k'^z + k^0 \, k'^0 + \un{k} \cdot
    \un{k}' \right] \, G_{22} (k, k') \notag \\ & = - \frac{1}{2} \int
  \frac{d^4 k}{(2 \pi)^4} \, \frac{d^4 k'}{(2 \pi)^4} \, e^{-i \,
    (k+k') \cdot x} \, \left[ k^+ \, k'^+ + k^- \, k'^- + \un{k} \cdot
    \un{k}' \right] \, G_{22} (k, k').
\end{align}
The saddle points now give (at $z=0$)
\begin{align}
  \left[ k^+ \, k'^+ + k^- \, k'^- + \un{k} \cdot \un{k}' \right]
  \bigg|_{z=0} = \left[ - \frac{k_T^2}{2} \, e^{\eta_1 - \eta_2} -
    \frac{k_T^2}{2} \, e^{\eta_2 - \eta_1} - k_T^2 \right]
  \bigg|_{\eta_1 = \eta_2 =0} = - 2 \, k_T^2,
\end{align}
such that the energy density is
\begin{align}
  \epsilon\big|_{\tau \gg \frac{1}{Q_s}} \approx - {\un{\nabla}}_1^2
  G_{22} (x_1, x_2) \Bigg|_{x_1 = x_2, \ \tau \gg \frac{1}{Q_s}} \sim
  \frac{1}{\tau},
\end{align}
again in agreement with the free-streaming behavior \eqref{eq:free}.

Finally, the transverse pressure is
\begin{align}
  \label{eq:pTresc}
  T^{11} & = \left\langle (\partial_x \varphi)^2 + {\cal L}
  \right\rangle \approx \left\langle \frac{1}{2} (\partial_t
    \varphi)^2 - \frac{1}{2} (\partial_z \varphi)^2 + \frac{1}{2}
    (\pd_x \varphi)^2 - \frac{1}{2} (\pd_y \varphi)^2 \right\rangle
  \notag \\ & = - \frac{1}{2} \int \frac{d^4 k}{(2 \pi)^4} \,
  \frac{d^4 k'}{(2 \pi)^4} \, e^{-i \, (k+k') \cdot x} \, \left[ k^0
    \, k'^0 - k^z \, k'^z \right] \, G_{22} (k, k') \notag \\ & = -
  \frac{1}{2} \int \frac{d^4 k}{(2 \pi)^4} \, \frac{d^4 k'}{(2 \pi)^4}
  \, e^{-i \, (k+k') \cdot x} \, \left[ k^+ \, k'^- + k^- \, k'^+
  \right] \, G_{22} (k, k').
\end{align}
Again, at the saddle points at mid-rapidity we get
\begin{align}
  \left[ k^+ \, k'^- + k^- \, k'^+ \right] \bigg|_{z=0} = - k_T^2 \,
  \cosh (\eta_1 - \eta_2) \bigg|_{\eta_1 = \eta_2 =0} = - k_T^2,
\end{align}
and the transverse pressure is
\begin{align}
  \label{eq:pTfin}
  P_T \big|_{\tau \gg \frac{1}{Q_s}} \approx - \frac{1}{2} \,
  {\un{\nabla}}_1^2 G_{22} (x_1, x_2) \Bigg|_{x_1 = x_2, \ \tau \gg
    \frac{1}{Q_s}} = \frac{1}{2} \, \epsilon\big|_{\tau \gg
    \frac{1}{Q_s}} \sim \frac{1}{\tau},  
\end{align}
in complete agreement with the free-streaming expectations
\eqref{eq:free}.


\section{Two-point correlation function with a single rescattering:
  Full diagrammatic calculation in the Wigner representation}
\label{sec:XP}

In this Section we calculate the late-time asymptotics of the
correlation function $G_{22}(X,p)$ in the Wigner representation. We
start with the calculation of the classical two-point correlation
function given by the diagram in Fig.~\ref{fig:g22cl}. Then, we
calculate $G_{22}(X,p)$ due to a single $2\to2$ rescattering using the
diagrams in Fig.~\ref{fig:phi4}.


\subsection{The asymptotic expansion of the classical correlation
  function in $1/\tau$}

\label{sec:Wigner_LO}

From (\ref{eq:Gcl_sc}) we have
\begin{align}
  G_{22}^{LO}(X,p)&=\int \frac{d^4k}{(2\pi)^4} e^{-ik \cdot X}G_{22}^{LO}(k/2+p, k/2-p)\nonumber\\
  &=-f(p_T)\int \frac{dk^+dk^-}{(2\pi)^2} e^{-ik^+ X^--ik^-
    X^+}G_R(k/2-p)G_R(k/2+p)
\end{align}
with $k_T=0$. One can integrate out $k^-$ by closing the integration
contour downward and picking up the residues of the poles in the lower
half-plane. There are two poles in the lower half-plane
\begin{align}
  \label{eq:k12p}
  k^-=k_{1p}^-\equiv \frac{2 p_\perp^2}{k^+-2 p^+}+2
  p^-\qquad\text{and}\qquad k^-=k_{2p}^-\equiv\frac{2 p_\perp^2}{k^++2
    p^+}-2 p^-,
\end{align}

and their residues yield
\begin{align}
  G_{22}^{LO}(X,p)&=-if(p_T)\int\frac{dk^+}{2\pi}\left.\frac{e^{-i (k^- X^++k^+ X^-)}}{{ p^-\left[\left(k^+\right)^2 - \frac{2 p^+ p^2}{p^-}\right]}}\right|_{k^-=k_{2p}^-}^{k^-=k_{1p}^-}\nonumber\\
  &=-if(p_T)\int\frac{dk^+}{2\pi}\left\{\frac{e^{-i \left[2 X^+ \left(\frac{p_T^2}{k^+-2 p^+}+p^-\right)+k^+ X^-\right]}}{{ p^-\left[\left(k^+\right)^2 - \frac{2 p^+ p^2}{p^-}\right]}}\right.\nonumber\\
  &\left.-\frac{e^{-i \left[\frac{2 X^+ p_T^2}{k^++2 p^+}+k^+ X^--2
          p^- X^+\right]}}{{ p^-\left[\left(k^+\right)^2 - \frac{2 p^+
            p^2}{p^-}\right]}}\right\}\label{eq:G22XpLO}
\end{align}
with the understanding that the $k^+$ integration contour is located
infinitesimally above the real axis.

\begin{figure}[htb]
\begin{center}
\includegraphics[width=0.7 \textwidth]{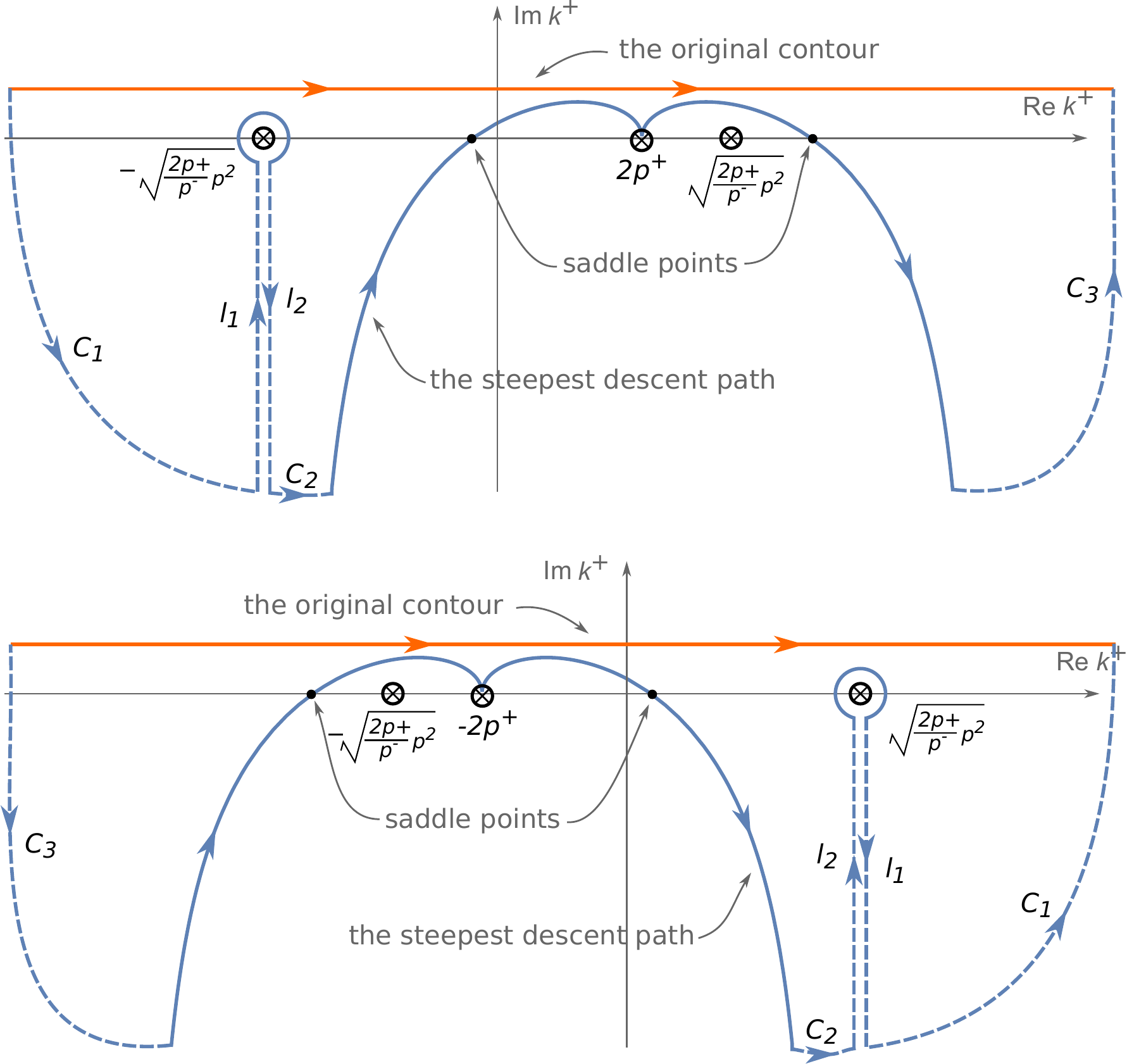}
\end{center}
\caption{Deformation of integration contours for the $1^{st}$ (top)
  and $2^{nd}$ (bottom) terms on the right-hand side of
  (\ref{eq:G22XpLO}). In each figure, the original integration contour
  is a horizontal line infinitesimally above the real axis. It is
  deformed so that it continues along the steepest descent path.  The
  contributions from integrating over $l_1$ and $l_2$ cancel with each
  other and the contributions from the arcs, $C_1$, $C_2$ and $C_3$,
  vanish as their radius goes to $\infty$.}
\label{fig:GXpLOContour}
\end{figure}

In this paper we are only interested in the asymptotic behavior of
$G_{22}(X,p)$ at large $\tau$. For this goal, we evaluate the two
terms on the right-hand side of (\ref{eq:G22XpLO}) separately. The
integrand of each term has a steepest descent path passing through two
saddle points respectively located at
\begin{align}
  k^+=2p^+\pm p_T\sqrt{\frac{2 X^+}{X^-}}\qquad\text{for the first
    term},\label{eq:sad1}
\end{align}
and
\begin{align}
  k^+=-2p^+\pm p_T\sqrt{\frac{2 X^+}{X^-}}\qquad\text{for the second
    term}.\label{eq:sad2}
\end{align}
As illustrated in Fig. \ref{fig:GXpLOContour}, we deform the
integration contours for these two integrals so that they continue
along the steepest descent paths. Each term has two other poles at
\begin{align}
  k^+=\pm 2 p^+ \sqrt{\frac{p^2}{2 p^+ p^-}}.\label{eq:poleLO}
\end{align}
Here, one only needs to consider the case with $\frac{p^2}{2 p^+
  p^-}>0$ since the residues of these poles will be exponentially
suppressed at large $\tau$ otherwise. The contour deformation may
involve passing through these poles. If this is the case, one needs to
pick up the contribution from their residues. By integrating over the
deformed contours one can easily pick up the contributions from the
residues of the poles (\ref{eq:poleLO}) and from the steepest descent
paths.

\begin{figure}[htb]
\begin{center}
\includegraphics[width=0.9\textwidth]{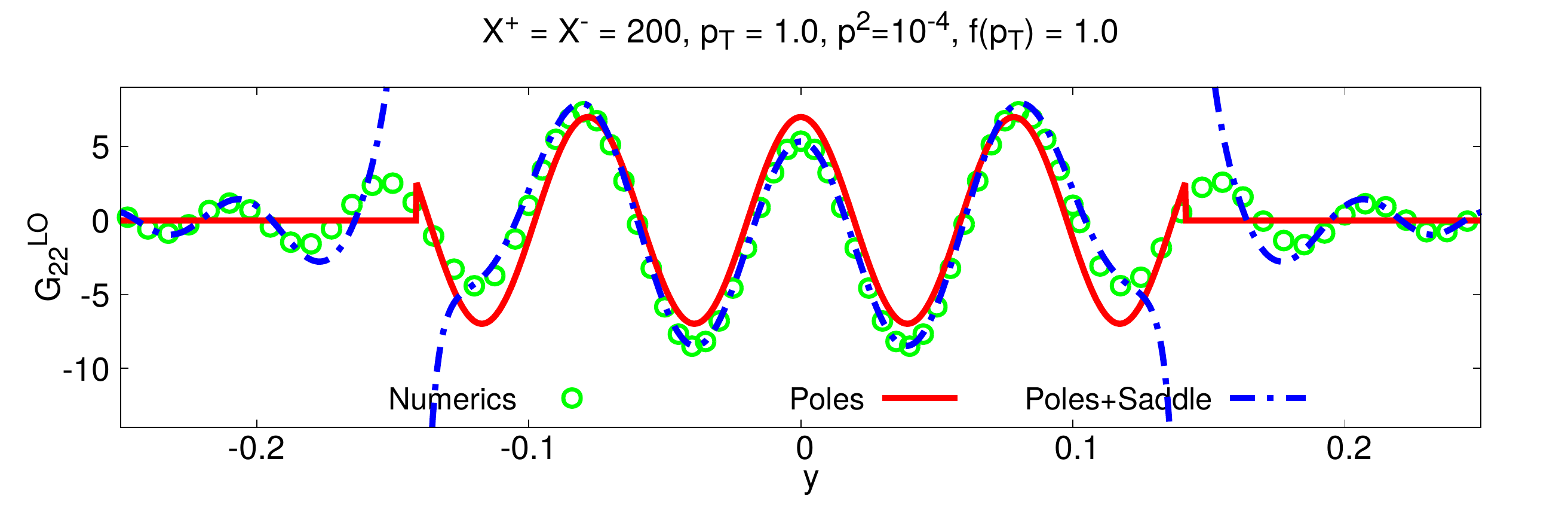}
\includegraphics[width=0.9\textwidth]{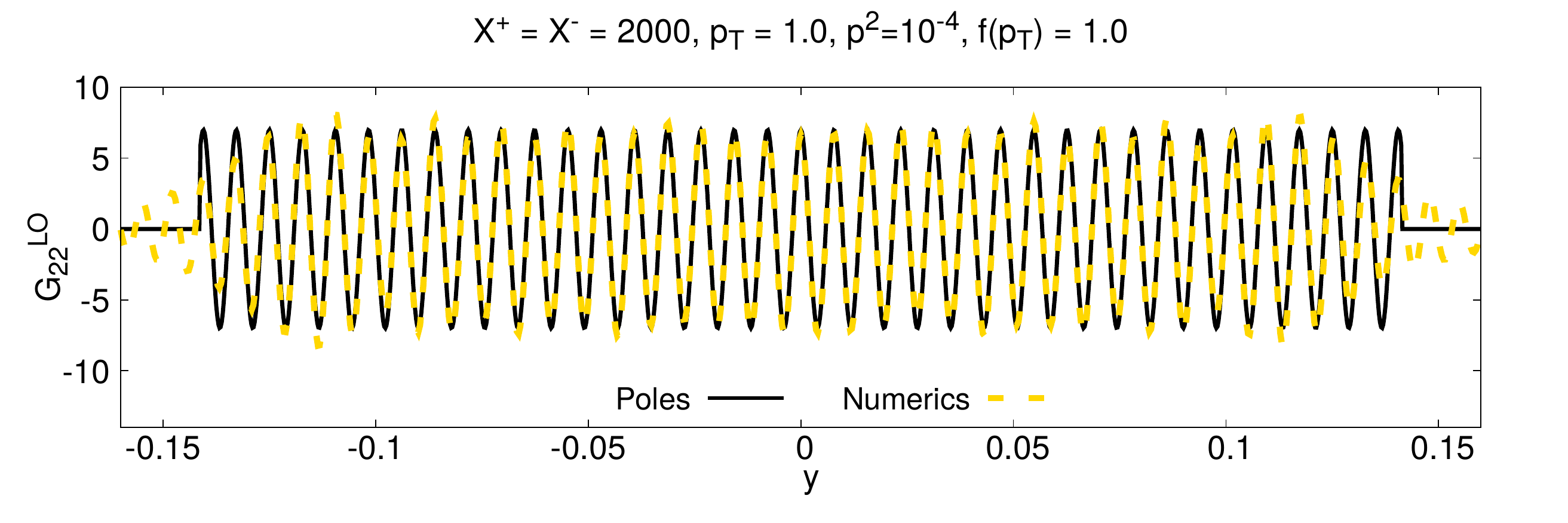}
\end{center}
\caption{Comparison with the numerical calculation of
  $G_{22}^{LO}(X,p)$. The top figure shows our analytical result in
  (\ref{eq:pole}) (Poles), the sum of (\ref{eq:pole}) and
  (\ref{eq:sadd}) (Poles+Saddle) and our numerical result (Numerics)
  at $X^+=X^-=200$. The bottom figure shows our result from poles
  (Poles) and our numerical result (Numerics) at $X^+=X^-=2000$. (The
  units are arbitrary.)}
\label{fig:GXpLONum}
\end{figure}

The residues of the poles yield a contribution with an amplitude of
$O(\tau^0)$ at large $\tau$. It is easy to see that the integrand on
the right hand side of (\ref{eq:G22XpLO}) as a whole is not singular
at the points in (\ref{eq:poleLO}). That is, the residues of the two
terms at the same pole cancel with each other. Therefore, we only need
to take into account the case when the deformed contours for the two
terms pass through different poles: $k^+=2p^+\sqrt{\frac{p^2}{2 p^+
    p^-}}$ for the first term and $k^+=-2p^+\sqrt{\frac{p^2}{2 p^+
    p^-}}$ for the second term. As illustrated in
Fig. \ref{fig:GXpLOContour}, for $p^+>0$ this requires
\begin{align}
  2p^++p_T\sqrt{\frac{2 X^+}{X^-}}>2p^+\sqrt{\frac{p^2}{2 p^+ p^-}} >
  2p^+-p_T\sqrt{\frac{2 X^+}{X^-}}>-2p^+\sqrt{\frac{p^2}{2 p^+ p^-}},
\end{align}
which is equivalent to
\begin{align}
  \tau>\frac{p_T}{2 |p^+ p^-|}|p^+X^-+p^-X^+|.\label{eq:taucond}
\end{align}
Similarly, for $p^+<0$ this requires
\begin{align}
  -2p^+ \sqrt{\frac{p^2}{2 p^+ p^-}} > 2p^++p_T\sqrt{\frac{2
      X^+}{X^-}}>2p^+\sqrt{\frac{p^2}{2 p^+ p^-}} >
  2p^+-p_T\sqrt{\frac{2 X^+}{X^-}},
\end{align}
which is also equivalent to (\ref{eq:taucond}). By combining the
results for both $p^+>0$ and $p^+<0$, we obtain the following
contribution to $G_{22}^{LO}(X,p)$ from the residues of the poles
\begin{align}
  \text{Poles}&=\frac{\cos \left(2\sqrt{\frac{p^2}{2 p^+ p^-} } \left(p^+ X^--p^- X^+\right)\right)}{2 p^- p^+ \sqrt{\frac{p^2}{2 p^- p^+}}}\nonumber\\
  &\times \theta\left(\frac{p^2}{2 p^- p^+}\right) \theta
  \left(\tau-\frac{p_T}{2|p^+ p^-|}\left| p^+ X^-+p^-X^+\right|
  \right) \, f(p_T).\label{eq:pole}
\end{align}
As shown in Fig. \ref{fig:GXpLONum}, the above result agrees very well
with our numerical result of $G_{22}^{LO}(X,p)$ at large $\tau$.

The integration over the steepest descent paths gives a contribution
with an amplitude of $O(\tau^{-\frac{1}{2}})$ at large $\tau$. Such a
contribution comes from the integration over a small region in the
vicinity of each saddle point in (\ref{eq:sad1}) and
(\ref{eq:sad2}). One only needs to expand the exponent of the
exponential function in a Taylor series around each saddle point up to
the second order and replace $k^+$ in the rest part of the integrand
by the saddle point value. By doing this and integrating out $k^+$, we
get
\begin{align}
  &\text{Saddle Points}=\frac{1}{2}\left(\frac{1}{2\pi p_T\tau}\right)^\frac{1}{2} \label{eq:sadd} \\
  &\times\left[\frac{\sin  \left(2 \left(p^- X^++p^+ X^--p_T\tau\right)\right)-\cos  \left(2 \left(p^- X^++p^+ X^--p_T\tau \right)\right)}{ \frac{p_T}{\tau} \left(p^- X^++p^+ X^-\right)-2 p^- p^+}\right.\nonumber\\
  &\left.\qquad+\frac{\sin \left(2 \left(p^- X^++p^+
          X^-+p_T\tau\right)\right)+\cos \left(2 \left(p^- X^++p^+
          X^-+p_T\tau\right)\right)}{ \frac{p_T}{\tau} \left(p^-
        X^++p^+ X^-\right)+2 p^- p^+}\right] \, f(p_T). \nonumber
\end{align}
As shown in the figure on the top of Fig.~\ref{fig:GXpLONum}, the
above result is indeed needed for us to understand the $\tau$
dependence of $G_{22}^{LO}(X,p)$ at an intermediate large time except
when the saddle points coincide with poles. This coincidence explains
the divergence between the exact numerical results and the dash-dotted
line in the top panel of \fig{fig:GXpLONum}. Our exact numerical
results in Fig.~\ref{fig:GXpLONum} show that $G_{22}^{LO}(X,p)$ is
well behaved in these regions. Hence, these regions should not be
important for calculating any observables involving $G_{22}(X,p)$ and
we do not need to construct a separate analytical expression for these
regions. As one can see from the lower panel of \fig{fig:GXpLONum},
these regions become progressively less important at later times.

In Ref. \cite{WuKovchegov} we find that the classical gluon two-point
function is
\begin{align}
  \left.G_{22}^{a\mu,b\nu}(X, p)\right|_{\tau\to\infty}\propto
  \delta^{ab} \sum\limits_{\lambda=\pm}\epsilon_\lambda^\mu(p)
  \epsilon_{\lambda}^{*\nu}(p)\frac{1}{\tau}\delta(p^2)\delta(y-\eta)
\end{align}
with
\begin{align}
  y\equiv\frac{1}{2}\ln\frac{p^+}{p^-},\qquad
  \eta\equiv\frac{1}{2}\ln\frac{X^+}{X^-}.
\end{align}
If we use
\begin{align}
  \lim\limits_{\tau\to \infty} \left[ \tau\frac{\cos(\tau y x)}{x}
  \right] = 2 \pi \delta(y)\delta(x^2),\label{eq:dd}
\end{align}
we obtain from \eq{eq:pole}
\begin{align}
  \left.G_{22}(X, p)\right|_{\tau\to\infty}=\frac{\pi}{\tau
    p_T}\delta\left(p^2\right)\delta(y-\eta)\theta\left(\frac{p^2}{2p^+p^-}\right)\theta(1-\cosh(y-\eta))
  \, f(p_T).
\end{align}
for our scalar correlator. In order to get the correct coefficient of
the above equation, we write
\begin{align}
  \left.G_{22}(X, p)\right|_{\tau\to\infty}=c_\theta\frac{\pi}{\tau
    p_T}\delta\left(p^2\right)\delta(y-\eta) \, f(p_T),
\end{align}
and fix $c_\theta$ by integrating out $p^\pm$ and matching it to that
by integrating (\ref{eq:pole}) over $p^\pm$. It is convenient to
define $\nu\equiv p^2=2 p^+ p^- - p_T^2$ and $y=\frac{1}{2}\ln
\left|\frac{p^+}{p^-}\right|$. In terms of these two variables,
Eq. (\ref{eq:pole}) reduces to, for $p^+ p^->0$,
\begin{align}\label{pole+} 
  \mbox{Poles} =\frac{\cos \left(2\tau\sqrt{{\nu}
      }\sinh(y-\eta)\right)}{\sqrt{\nu (\nu+p_T^2)}}\theta \left( \nu
    -p_T^2\sinh^2(y-\eta) \right) \, f(p_T),
\end{align}
and, for $p^+ p^-<0$,
\begin{align}\label{pole-} 
  \mbox{Poles}=-\frac{\cos \left(2\tau\sqrt{{-\nu}
      }\cosh(y-\eta)\right)}{\sqrt{\nu (\nu+p_T^2)}} \theta \left(-\nu
    -p_T^2\cosh^2(y-\eta) \right) \, f(p_T).
\end{align}
Accordingly, we have
\begin{align}
  \int\limits_{-\infty}^\infty dp^+ dp^- \mbox{Poles}=
  \left(\int\limits_{-k_T^2}^{+\infty} d\nu
    \int\limits_{-\infty}^{+\infty} dy+\int\limits_{-\infty}^{-k_T^2}
    d\nu \int\limits_{-\infty}^{+\infty} dy\right)\mbox{Poles}\equiv
  n_1+n_2
\end{align}
with only \eq{pole+} contributing to $n_1$ and only \eq{pole-}
contributing to $n_2$. After changing variables to
$\hat{s}=\sinh(y-\eta)$ and $\hat{\nu}=\sqrt{p^2}/p_T$, we have
\begin{align}
  n_1=2 \int\limits_0^\infty d\hat\nu \int\limits_0^\infty d\hat{s}
  \frac{\cos(2\tau p_T \hat{\nu}
    \hat{s})}{\sqrt{(\hat{\nu}^2+1)(\hat{s}^2+1)}} \, f(p_T).
\end{align}
We only need to obtain the large $\tau$ asymptotics. At large $\tau$,
the predominant contribution to the above integral comes from the
region $\hat{\nu}\approx 0$ and $\hat{s}\approx 0$. So one can not
take either $\tau\hat{\nu}$ or $\tau \hat{s}$ as the large expansion
parameter. We do the following trick: we rotate the integration
contour of, say $\hat s$, to go along the positive imaginary
axis. Then the $[0,i]$ region of integration does not contribute to
the real part of the integral. We get
\begin{align}
  n_1&=2 \text{Re} \int\limits_0^\infty d\hat \nu \int\limits_0^\infty
  d\hat s \frac{e^{i2\tau p_T \hat\nu \hat s}}{\sqrt{(\hat
      \nu^2+1)(\hat s^2+1)}} \, f(p_T) =
  2 \int\limits_0^\infty d\hat \nu \int\limits_1^\infty d\hat s \frac{e^{-2\tau p_T \hat \nu \hat s}}{\sqrt{(\hat \nu^2+1)(\hat s^2-1)}} \, f(p_T) \nonumber\\
  &\approx 2 \int\limits_1^\infty \frac{d\hat s}{2\tau p_T \hat s
    \sqrt{\hat s^2-1}} \, f(p_T) = \frac{\pi}{2 \, \tau \, p_T} \,
  f(p_T).
\end{align}
Similarly, by changing variables to $\hat s=\cosh(y-\eta)$ and $\hat
\nu=\sqrt{-p^2}/p_T$, we have
\begin{align}
  n_2&=-2 \! \int\limits_1^\infty d\hat \nu \int\limits_1^\infty d\hat s \frac{\cos(2\tau p_T \hat \nu \hat s)}{\sqrt{(\hat \nu^2-1)(\hat s^2-1)}} \, f(p_T) \approx-2 \text{Re} \! \int\limits_1^\infty d\hat \nu \frac{e^{i2\tau p_T \hat \nu-i\frac{\pi}{4}}}{\sqrt{(\hat \nu^2-1)}}\frac{i}{2}\left(\frac{\pi}{\tau p_T\hat \nu}\right)^\frac{1}{2} \! f(p_T)\nonumber\\
  &\approx\frac{\pi}{2 \tau p_T} \sin (2\tau p_T) \, f(p_T).
\end{align}
One can simply discard $n_2$ since it is a highly oscillatory function at large $\tau$. From $n_1$ we obtain $c_\theta = 1/2$ and
\begin{align}
  \label{eq:GXpLOasy}
\left.G_{22}(X,p)\right|_{\tau\to\infty}= \frac{\pi}{2 \tau p_T}\delta(y-\eta)\delta(p^2) \, f(p_T).
\end{align}


\subsection{$G_{22}(X,p)$ from a singe rescattering}

Let us evaluate the late time asymptotics of each diagram in
\fig{fig:phi4} using the same techniques as for $G_{22}^{LO}(X,p)$ in
the previous subsection.


\subsubsection{Diagram I}

From (\ref{eq:DiagIk2}), we have in Wigner representation
\begin{align}
  \label{eq:DiagIGXp1}
  \mbox{Diagram I} = & - \frac{\lambda^2}{2}\int\frac{d^2 k_1}{(2 \pi)^2} \, f(k_{1T}) \, f(k_{2T})\int\frac{d^4 k_3}{(2 \pi)^4} \, \pi\, \delta (k_3^2)\int\limits_{-\infty}^\infty \frac{d k^+ \, d k^-}{(2 \pi)^2}\notag\\
  &\times \, e^{- i k^+ X^- - i k^- X^+}
  \,G_R\left(\frac{k}{2}+p\right)G_R\left(\frac{k}{2}-p\right)\Pi(k^+,k^-),
\end{align}
where
\begin{align}
  \Pi(k^+,k^-) &\equiv  I_1 \left(\frac{k^+}{2}+p^+ - k_3^+, \frac{k^-}{2}+p^- - k_3^-, k_{1T}, k_{2T}\right) \, \notag\\
  &\times I_1 \left(\frac{k^+}{2}-p^+ + k_3^+, \frac{k^-}{2}-p^- +
    k_3^-, k_{1T}, k_{2T}\right)
\end{align}
with $\un{k}=0$ and $\un{k}_2=\un{p}-\un{k}_1-\un{k}_3$. The
integration contours for $k^+$ and $k^-$ are understood to be located
infinitesimally above the real axis of the $k^\pm$ plane and we shall
drop all the $i\epsilon$'s in both $G_R$ and $I_1$.

\begin{figure}[htb]
\begin{center}
\includegraphics[width=0.7\textwidth]{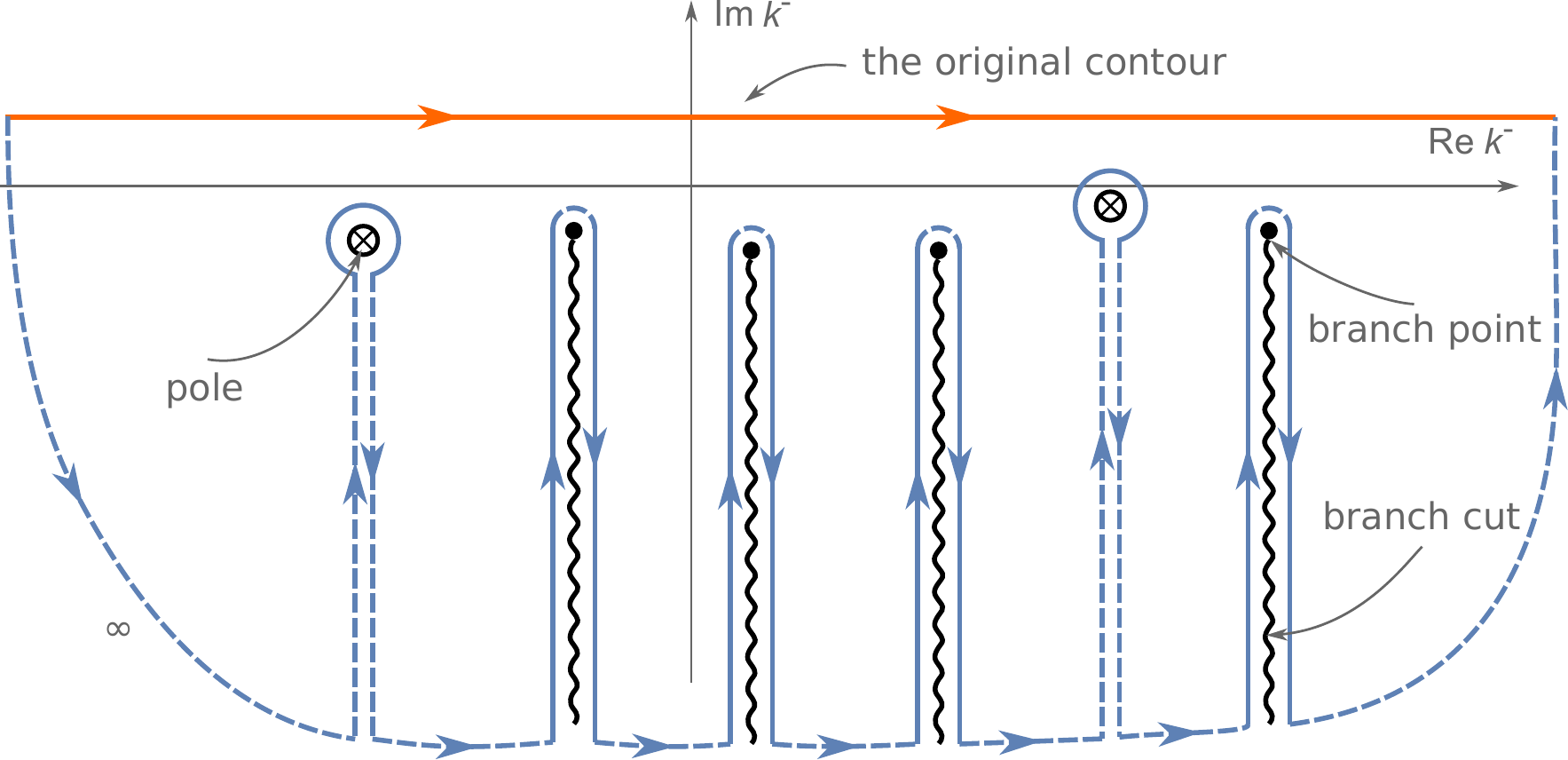}
\end{center}
\caption{Integration contour for $k^-$. In the integration over $k^-$
  in (\ref{eq:DiagIGXp1}), the integration contour is deformed in
  order to pick up the predominant contribution at late times from the
  regions around singular points (poles and branch points).  The
  portions of the contour giving vanishing or canceling contributions
  are indicated by the dashed lines.}
\label{fig:GXpNLOkm}
\end{figure}

In order to calculate the late-time asymptotics, we first need to
identify all the critical points including poles, branch points and
saddle points. For $k^-$, the integrand on the right hand of
(\ref{eq:DiagIGXp1}) has two poles given by (\ref{eq:k12p}) and 4
branch points at
\begin{align}
  k^-= 2 \left[\frac{\left(k_{1 T}\pm k_{2 T}\right){}^2}{k^++2
      k_3^+-2 p^+}-k_3^-+p^-\right]
\end{align}
and
\begin{align}
  k^-= 2 \left[\frac{\left(k_{1 T}\pm k_{2 T}\right){}^2}{k^+-2
      k_3^++2 p^+}+k_3^--p^-\right].
\end{align}
As shown in Fig. \ref{fig:GXpNLOkm}, we deform the integration contour
such that it goes around the two poles and continues along the branch
cuts. Then, $k^-$ integration is given by the residues of the two
poles and the integration along the branch cuts, as indicated by the
solid lines in Fig. \ref{fig:GXpNLOkm}. At late times, the poles give
a contribution proportional to $\tau^0$ while the branch cuts yield a
contribution proportional to $\tau^{-\frac{1}{2}}$. Hence, we only
keep the contribution from the residues of the poles. As a result, we
get
\begin{align}
  \label{eq:DiagIGXp2}
  \mbox{Diagram I} = & -\frac{\lambda^2}{2p^-}\int\frac{d^2 k_1}{(2 \pi)^2} \, f(k_{1T}) \, f(k_{2T})\int\frac{d^4 k_3}{(2 \pi)^4} \, \pi\, \delta (k_3^2)\int\limits_{-\infty}^\infty \frac{d k^+ \, }{2 \pi}\notag\\
  &\times \,\frac{i}{\left(k^+\right)^2 -\frac{2 p^+}{p^-} p^2}
  \left[\left. e^{- i k^+ X^- - i k^- X^+} \,\Pi(k^+,
      k^-)\right]\right|^{k^-= \frac{2 p_T^2}{k^+-2 p^+}+2 p^-}_{k^-=
    \frac{2 p_T^2}{k^++2 p^+}-2 p^-}.
\end{align}

For $k^+$, the integrand on the right hand side of
(\ref{eq:DiagIGXp2}) has poles given by (\ref{eq:poleLO}). It also has
saddle points and branch cuts. Since they only give a contribution
proportional to $\tau^{-\frac{1}{2}}$, we only need to keep the
contribution from the residues of the poles. The calculation is
straightforward and similar to the leading-order case of
Sec.~\ref{sec:Wigner_LO}. We obtain
\begin{align}
  \label{eq:DiagIGXp3}
  \mbox{Diagram I} = & \frac{\lambda^2}{4p^+p^-\sqrt{\frac{p^2}{2 p^+ p^-}}}~\text{Re} \Bigg[ \int\frac{d^2 k_1}{(2 \pi)^2} \, f(k_{1T}) \, f(k_{2T})\int\frac{d^4 k_3}{(2 \pi)^4} \, \pi\, \delta (k_3^2) \notag\\
  &\times\theta\left(\frac{p^2}{2 p^- p^+}\right)  \theta \left(\tau-\frac{p_T}{2|p^+ p^-|}\left| p^+ X^-+p^-X^+\right| \right) \,\notag\\
  &\times e^{2 i \sqrt{\frac{p^2}{2 p^+ p^-}}\left( p^+ X^- - p^-
      X^+\right)} \Pi \left(-2 p^+\sqrt{\frac{p^2}{2 p^+ p^-}}, 2
    p^-\sqrt{\frac{p^2}{2 p^+ p^-}}\right) \Bigg] \,,
\end{align}
where we have used the identity
\begin{align}\label{II*}
  I_1(-q^+,-q^-,k_{1T},k_{2T})=I_1^*(q^+,q^-,k_{1T},k_{2T})
\end{align}
to shorten the expression by taking the real part. Based on the same
calculation as for (\ref{eq:GXpLOasy}) we obtain
\begin{align}
  \label{eq:DiagIGXp3}
  \left.\mbox{Diagram I}\right|_{\tau\to\infty}&= \frac{\lambda^2}{2}\int\frac{d^2 k_1}{(2 \pi)^2} \, f(k_{1T}) \, f(k_{2T})\int\frac{d^4 k_3}{(2 \pi)^4} \, \pi\, \delta (k_3^2)\,\notag\\
  &\times \frac{\pi}{2\, \tau \,
    p_T}\delta(y-\eta)\delta(p^2)\left|I_1 \left(p^+ - k_3^+, p^- -
      k_3^-, k_{1T}, k_{2T}\right)\right|^2.
\end{align}


\subsubsection{Diagrams II and II'}

\begin{figure}[htb]
\begin{center}
\includegraphics[width=\textwidth]{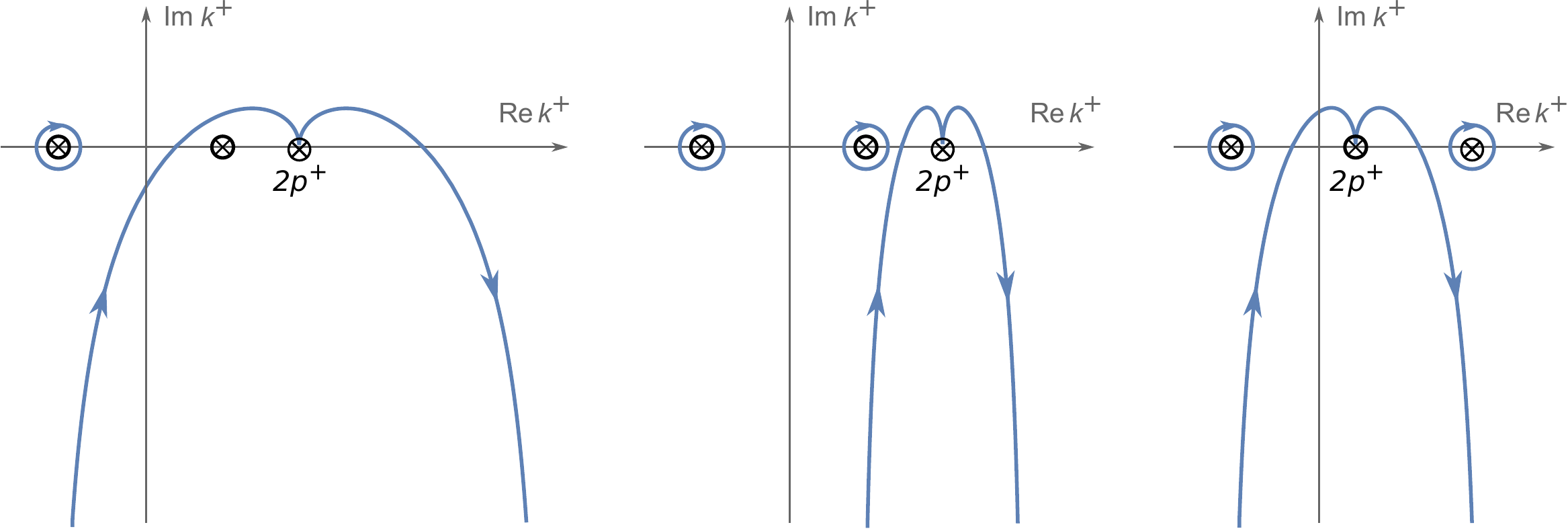}
\end{center}
\caption{Relative positions of poles with respect to the steepest descent path. Here, we assume that $p^+>0$ and that the two poles are located at $k^+=\pm 2p^+\sqrt{\frac{p^2}{2p^+ p^-}}$. These 3 figures show all the possible cases when one needs to keep the residues of the poles in the $k^+$ integration in (\ref{eq:DiagIIGXp2}).}
\label{fig:GXpDiagII}
\end{figure}

From (\ref{eq:DiagIIk}) we have in Wigner representation
\begin{align}
  \label{eq:DiagIIGXp1}
  \mbox{Diagrams II+II'} = & - \frac{\lambda^2}{2}\int\frac{d^2
    k_1}{(2 \pi)^2} \, f(k_{1T}) \, f(k_{2T})\int\frac{d^4 k_3}{(2
    \pi)^4} \int\limits_{-\infty}^\infty \frac{d k^+ \, d k^-}{(2
    \pi)^2} \,
  e^{- i k^+ X^- - i k^- X^+} \,\notag\\
  &\times\left[\,\pi \delta\left(\left(\frac{k}{2}+p\right)^2\right)G_R\left(\frac{k}{2}-p\right) G_A(k_3)\right.\,\notag\\
  &+\left.\,G_R\left(\frac{k}{2}+p\right) \pi
    \delta\left(\left(\frac{k}{2}-p\right)^2\right)
    G_R(k_3)\right]\Pi(k^+, k^-),
\end{align}
with $\un{k}=0$ and $\un{k}_2=\un{p}-\un{k}_1-\un{k}_3$. By
integrating out $k^-$ we obtain
\begin{align}
  \label{eq:DiagIIGXp2}
  \mbox{Diagrams II+II'} =&  - \frac{\lambda^2}{4p^-}\int\frac{d^2 k_1}{(2 \pi)^2} \, f(k_{1T}) \, f(k_{2T})\int\frac{d^4 k_3}{(2 \pi)^4} \int\limits_{-\infty}^\infty \frac{d k^+ }{2 \pi} \frac{i}{\left(k^+\right)^2 -\frac{2 p^+}{p^-} p^2}\,\notag\\
  &\times\left[\,e^{- i k^+ X^- - i k_{1p}^- X^+}~\mbox{Sign}\left(k^+-2p^+\right) G_R(k_3)\Pi(k^+, k_{1p}^-)\right.\,\notag\\
  &-\left.e^{- i k^+ X^- - i k_{2p}^- X^+}~\mbox{Sign}\left(k^++2
      p^+\right) G_A(k_3)\Pi(k^+, k_{2p}^-)\right]
\end{align}
with $k_{1p}^-$ and $k_{2p}^-$ given in (\ref{eq:k12p}). By using
\eq{II*} one can show that the second term on the right hand side of
(\ref{eq:DiagIIGXp2}) gives a contribution which is the complex
conjugate of the first term.

We only evaluate the late time asymptotics from diagrams II and
II'. The integrand on the right hand side of (\ref{eq:DiagIIGXp2}) has
poles, saddle points and branch points. As before, we deform the
integration contour so that it wraps around poles, continues along the
steepest descent path and goes along the branch cuts. Then, at large
$\tau$ we only need to pick up the contribution from the poles. Since
there is no obvious cancellation between the residues of the poles for
the two terms in (\ref{eq:DiagIIGXp2}), we need to keep the
contributions from all the poles which are in the way of the contour
deformation. As shown in Fig. \ref{fig:GXpDiagII}, there are 3
possible cases when we need to keep the residues of
poles. Correspondingly, they yield
\begin{align}
  \label{eq:DiagIIGXp3}
  \mbox{Diagrams II+II'} & =  - \frac{\lambda^2}{8p^+p^-\sqrt{\frac{p^2}{2 p^+ p^-}}}\theta\left(\frac{p^2}{2 p^- p^+}\right) \text{Sign}(p^+)\int\frac{d^2 k_1}{(2 \pi)^2}\frac{d^4 k_3}{(2 \pi)^4} \, f(k_{1T}) \, f(k_{2T})\notag\\
  &\times~\text{Re}~\Bigg\{ G_R(k_3) \,\left[\, \theta \left(\tau-\frac{p_T}{2|p^+ p^-|}\left| p^+ X^-+p^-X^+\right| \right)\right.\notag\\
  &\left.+ 2 i \, \theta\left( \frac{2|p^+|}{p_T}\left(1-\sqrt{\frac{p^2}{2p^+ p^-}}\right)-\sqrt\frac{2X^+}{X^-} \right)\text{Im}\right.\,\notag\\
  &\left.+ 2 \, \theta\left( \frac{2|p^+|}{p_T}\left(\sqrt{\frac{p^2}{2p^+ p^-}}-1\right)-\sqrt\frac{2X^+}{X^-} \right)\text{Re}\right] \,\notag\\
  &\times e^{2 i \sqrt{\frac{p^2}{2 p^+ p^-}}\left( p^+ X^- - p^-
      X^+\right)} \Pi \left(-2p^+\sqrt{\frac{p^2}{2 p^+ p^-}},
    2p^-\sqrt{\frac{p^2}{2 p^+ p^-}}\right) \Bigg\},
\end{align}
where Re and Im in the curly brackets act on everything to their
right. Taking the large-$\tau$ limit using \eq{eq:dd} we see that only
the cosine part of the exponential contributes, putting $y=\eta$ and
$p^2 =0$ while approaching this latter limit from the side where
$p^2/(2 p^+ p^-) >0$. This way, the second and third terms in the
square brackets vanish. To take the real part in the first term we
employ the Cutkosky rules \cite{Cutkosky:1960sp}, which prescribe the
replacement
\begin{align}
  \label{eq:repl_Cut}
  \frac{1}{k_3^2 + i \epsilon k_3^+} \to - 2 \pi i \, \delta (k_3^2)
  \, \mbox{Sign} (k_3^+).
\end{align}
This gives
\begin{align}
  \label{eq:DiagIIGXp}
  &\left.\mbox{Diagrams II+II'}\right|_{\tau\to\infty} =  - \frac{\lambda^2}{2}\text{Sign}(p^+)\int\frac{d^2 k_1}{(2 \pi)^2}\frac{d^4 k_3}{(2 \pi)^4} \, f(k_{1T}) \, f(k_{2T})\notag\\
  &\times~\text{Sign}(k_3^+)\pi\delta(k_3^2)\frac{\pi}{2\, \tau \,
    p_T}\delta(y-\eta)\delta(p^2)\left|I_1 \left(p^+ - k_3^+, p^- -
      k_3^-, k_{1T}, k_{2T}\right)\right|^2.
\end{align}
Adding this result to diagram I leads to a cancellation similar to
that observed in \eq{eq:simpl}. In the end we obtain
\begin{align}
  \label{eq:Gain}
  &\left.\mbox{Diagrams I+ II+II'}\right|_{\tau\to\infty} =  \frac{\pi\lambda^2}{2\tau}\delta(y-\eta)\delta(p^2)\int\frac{d^2 k_1}{(2 \pi)^2}\frac{d^4 k_3}{(2 \pi)^4} \, f(k_{1T}) \, f(k_{2T})\notag\\
  &\qquad\times~\theta(-k_3^+ p^+) \, \pi\delta(k_3^2)\left|I_1
    \left(p^+ - k_3^+, p^- - k_3^-, k_{1T}, k_{2T}\right)\right|^2.
\end{align}


\subsubsection{Diagrams III and III'}

Let us now evaluate diagrams III and III' from
Fig. \ref{fig:phi4}. Similar to that in coordinate space, we have, now
in Wigner representation
\begin{align}
  &\mbox{Diagrams III+III'}=-\lambda^2 \int\limits_{-\infty}^\infty \frac{d k^+ d k^- }{(2 \pi)^2}e^{- i k^+ X^- - i k^- X^+}G_R\left(\frac{k}{2}+p\right)G_R\left(\frac{k}{2}-p\right)\notag\\
  &\times\int\frac{d^4k_1}{(2 \pi)^4}\frac{d^4k_3}{(2 \pi)^4} \, f(p_{T}) \, f(k_{2T})\left[G_R\left(\frac{k}{2}+p-k_1-k_3\right)+G_R\left(\frac{k}{2}-p-k_1-k_3\right)\right]\notag\\
  &\times G_R(k_1)\pi\delta(k_3^2) I_1(k_1^++k_3^+,k_1^-+k_3^-,
  k_{2T}, p_T).
\end{align}
It is easy to see that diagram III' is a complex conjugate of diagram
III. By employing this fact and (\ref{eq:identity}) at large $\tau$,
we can write
\begin{align}
  &\mbox{Diagrams III+III'}=-\frac{2 \lambda^2}{p^-} \mbox{Re}\int\frac{d^2k_1}{(2 \pi)^2}\frac{d^2k_3}{(2 \pi)^2}\, f(p_{T}) \, f(k_{2T})\frac{d q^+ dq^-}{(2 \pi)^2}I_1(q^+, q^-, k_{2T}, p_T)\notag\\
  &\times \left. \int\limits_{-\infty}^\infty \frac{d k^+}{2 \pi} \frac{ie^{- i k^+ X^- - i k^- X^+}}{(k^+)^2-\frac{2p^+}{p^-} p^2}I_4(q^+, q^-, k_{3T},k_{1T})G_R\left(\frac{k}{2}+p-q\right)\right|^{k^-=k^-_{1p}}_{k^-=k^-_{2p}}\notag\\
  &=-\frac{2 \lambda^2}{p^-} \mbox{Re}\int\frac{d^2k_1}{(2 \pi)^2}\frac{d^2k_3}{(2 \pi)^2}\, f(p_{T}) \, f(k_{2T}) \int\limits_{-\infty}^\infty \frac{d k^+}{2 \pi} \frac{i}{(k^+)^2-\frac{2p^+}{p^-} p^2}\notag\\
  &\times \left[\left.e^{- i k^+ X^- - i k^-
        X^+}J\left(\frac{k^+}{2}-p^+,p_T,k_{1T},k_{2T},k_{3T}\right)\right]\right|^{k^-=k^-_{1p}}_{k^-=k^-_{2p}} ,
\end{align}
where we put $\un{q} = \un{k}_1 + \un{k}_3$.  Using (\ref{eq:Jsimpl})
leads to
\begin{align}
  &\mbox{Diagrams III+III'}=-\frac{2 \lambda^2}{p^-} \mbox{Re}\int\frac{d^2k_1}{(2 \pi)^2}\frac{d^2k_3}{(2 \pi)^2}\, f(p_{T}) \, f(k_{2T}) \int\limits_{-\infty}^\infty \frac{d k^+}{2 \pi} \frac{i}{(k^+)^2-\frac{2p^+}{p^-} p^2}\notag\\
  &\times \left[\left.e^{- i k^+ X^- - i k^- X^+}J\left(p_T,k_{1T},k_{2T},k_{3T}\right)\right]\right|^{k^-=k^-_{1p}}_{k^-=k^-_{2p}}.
\end{align}
At the end we have
\begin{align}
  &\mbox{Diagrams III+III'}=\frac{2 \lambda^2}{p^-}
  \mbox{Re}\int\frac{d^2k_1}{(2 \pi)^2}\frac{d^2k_3}{(2 \pi)^2}\,
  f(p_{T}) \, f(k_{2T})J\left(p_T,k_{1T},k_{2T},k_{3T}\right)
  \theta\left(\frac{p^2}{2 p^- p^+}\right)
  \notag\\
  &\times \theta \left(\tau-\frac{p_T}{2|p^+ p^-|}\left| p^+
      X^-+p^-X^+\right| \right)\frac{\cos \left(2\sqrt{\frac{p^2}{2
          p^+ p^-} } \left(p^+ X^--p^- X^+\right)\right)}{2 p^- p^+
    \sqrt{\frac{p^2}{2 p^- p^+}}}.
\end{align}
That is, at late times
\begin{align}
  \label{eq:GXpDiagIII}
  \mbox{Diagrams III+III'} \Bigg|_{\tau \to \infty}
  \sim \frac{1}{\tau}\delta(p^2)\delta(\eta-y).
\end{align}

Combining \eqref{eq:Gain} and \eqref{eq:GXpDiagIII} we conclude once
again that the late-time asymptotics of the rescattering diagrams in
\fig{fig:phi4}, calculated this time in Wigner representation, is
consistent with free streaming.


\section{Conclusions and Outlook}
\label{sec:disc}

In the first paper \cite{WuKovchegov} of this `duplex' we have
outlined the way to apply the Schwinger--Keldysh formalism to
ultrarelativistic heavy ion collisions, thus setting the stage for the
calculation of time-dependent observables, such as the energy-momentum
tensor, in the collisions. As the first application of this technique,
in \cite{WuKovchegov} we have tried to re-derive the Boltzmann
equation for the medium produced in heavy ion collisions by
considering a single $2\to 2$ rescattering correction to the classical
gluon fields of the MV model. We have employed the ``on-shell''
approximation for the propagators that one usually employs in deriving
the kinetic theory. The result was dependent on the time-ordering
assumptions outlined above in Sec.~\ref{sec:intro}: kinetic theory
emerged under the assumption (i), while free-streaming was obtained if
assumption (ii) was employed.

In this paper we have used the formalism set up in \cite{WuKovchegov}
to re-do the calculation of the $2\to 2$ rescattering correction to
the classical fields without using the ``on-shell'' approximation for
the propagators and without assuming a specific time-ordering of the
interaction time versus the measurement time. Performing the
calculation twice, both in momentum space (Sec.~\ref{sec:x1x2}) and in
the Wigner representation (Sec.~\ref{sec:XP}), we have arrived at the
results consistent with free streaming \eqref{eq:free}. We have thus
found no evidence for the applicability of the kinetic theory
(employing the Boltzmann equation taken with the full collision term)
to the perturbative description of heavy ion collisions.

Our perturbative calculations of the energy-momentum tensor are
consistent with the conclusion reached in \cite{Kovchegov:2005ss},
where it was argued that the energy density of the produced
weakly-coupled medium is given by
\begin{align}
  \label{eq:energy_density}
  \epsilon (\tau, \eta, \un{b}) \bigg|_{\tau \gg 1/Q_s} \, \approx \,
  \frac{1}{\tau} \, \int d^2 k_T \ k_T \ \frac{d N}{d^2 k_T \, d \eta
    \, d^2 b_T}
\end{align}
at any order in perturbation theory. In the scenario advocated in
\cite{Kovchegov:2005ss}, higher-order perturbative corrections would
only modify the gluon multiplicity distribution $d N/d^2 k_T \, d \eta
\, d^2 b_T$ and the saturation scale $Q_s$, leaving the
$\tau$-dependence of \eq{eq:energy_density} unchanged.

In the future we hope that the formalism we have presented in
\cite{WuKovchegov} will be useful for calculations of time-dependent
heavy ion observables. In particular it can be used to cross-check
(and possibly challenge) the conclusion in \eqref{eq:energy_density}
by explicit perturbative calculations. Our own cross-check presented
above did not show any deviations from \eqref{eq:energy_density} and
disagreed with kinetic theory. Perhaps other thermalization scenarios
may fare better in challenging \eq{eq:energy_density}. Indeed
calculations of higher-order perturbative corrections to the classical
gluon field contributions to heavy ion observables appear to be very
complicated. In our minds, however, such calculations would present a
necessary theoretical test for any thermalization scenario. If, for
instance, a given thermalization proposal claims to resum a certain
$\tau$-dependent parameter to all orders, then this parameter should
manifest itself in some lower-order perturbative calculation. In other
words, one needs to prove that the resummation parameter exists. If
the explicit calculation fails to confirm that the resummation
parameter exists, the thermalization scenario in question should be
discarded. If the equation \eqref{eq:energy_density} does turn out to
be exact in the perturbation theory, then calculation of higher-order
corrections would improve our knowledge of gluon multiplicity and
energy density, which would also be very useful.  We hope the
formalism and calculations presented in \cite{WuKovchegov} and in this
paper lay the groundwork for future cross-checks of thermalization
scenarios and will facilitate calculations of heavy-ion observables in
the years to come.


\acknowledgments

The authors would like to thank Mauricio Martinez for the extensive
discussions of thermalization in heavy ion collisions which got YK
interested in this project. We also thank Hong Zhang and Gojko
Vujanovic for discussions and advice. This material is based upon work
supported by the U.S. Department of Energy, Office of Science, Office
of Nuclear Physics under Award Number DE-SC0004286.


\appendix

\section{$I_1$ Calculation}
\label{sec:I1}

The goal of this Appendix is to calculate
\begin{align}\label{I11}
  I_1(q^+,q^-,k_T,p_T)=\int\limits_{-\infty}^\infty \frac{dk^+
    dk^-}{(2\pi)^2} \, G_R(k) \, G_R(q-k)
\end{align}
with
\begin{align} 
  {\un p} = {\un q} - {\un k}.
\end{align}
Taking $G_R$ from \eq{eq:Green_ret} and integrating \eqref{I11} over
$k^-$ we get
\begin{align}
  \label{eq:I12}
  I_1 =\int\limits_{-\infty}^\infty \frac{dk^+}{2\pi} \frac{i}{2 (q^+
    - k^+) (2 q^- k^+ - k_T^2) - 2 k^+ \, p_T^2 + i \, \epsilon \, k^+
    (q^+ - k^+)}.
\end{align}
The poles of the integrand are given by
\begin{align}
  \label{eq:k12}
  k^{+}_{1,2} = \frac{1}{- 4 q^-} \left[ p_T^2 - k_T^2 - 2 q^+ q^- \pm
    \sqrt{D} \right]
\end{align}
with the discriminant 
\begin{align}
  \label{eq:D}
  D & \equiv [(k_T-p_T)^2-2 q^+ q^-] [(k_T+p_T)^2-2 q^+ q^-] \notag \\
  & = (k_T^2+p_T^2- 2q^+ q^-)^2- 4 \, k_T^2 \, p_T^2.
\end{align}
Using $k^{+}_{1,2}$ we rewrite \eq{eq:I12} as
\begin{align}
  \label{eq:I13}
  I_1 = \frac{1}{- 4 q^-} \int\limits_{-\infty}^\infty
  \frac{dk^+}{2\pi} \frac{i}{(k^+ - k_1^+ + i \epsilon r_1) \, (k^+ -
    k_2^+ + i \epsilon r_2)},
\end{align}
where the coefficients $r_1, r_2$ have to be determined by matching
the linear in $\epsilon$ terms in the denominators of \eqref{eq:I13}
and \eqref{eq:I12} at the poles $k^+ = k_1^+$ and $k^+ = k_2^+$
respectively. This gives
\begin{subequations}  \label{eq:r12}
  \begin{align}
  & - 4 q^- (k_1^+ - k_2^+) \, r_1 = k_1^+ \, (q^+ - k_1^+),  \label{eq:r12a} \\
  & - 4 q^- (k_2^+ - k_1^+) \, r_2 = k_2^+ \, (q^+ - k_2^+).
\end{align}
\end{subequations}
Multiplying Eqs.~\eqref{eq:r12} after a little algebra we arrive at
\begin{align}
  \label{eq:r12sign}
  - (4 q^-)^2 \, (k_1^+ - k_2^+)^2 \, r_1 \, r_2 = k_T^2 \, p_T^2 \,
  \left( \frac{q^+}{2 \, q^-} \right)^2.
\end{align}
We are now ready to integrate \eq{eq:I13} over $k^+$. To do so we need
to consider two cases:
\begin{itemize}
\item {\bf Case I:} $D<0$. In this case $k_1^+$ and $k_2^+$ have
  non-zero (and non-infinitesimal) imaginary parts, and the $i
  \epsilon$ terms along with the values of $r_1$ and $r_2$ in
  \eq{eq:I13} are not important. Integrating \eq{eq:I13} over $k^+$ we
  obtain
  \begin{align}
    \label{eq:I14}
    I_1 \Big|_{D<0} = - \frac{i}{2 \sqrt{|D|}} \, \mbox{Sign} (q^-) =
    - \frac{i}{2 \sqrt{|D|}} \, \mbox{Sign} (q^+).
  \end{align}
  (We have used the fact that $q^+ q^- >0$ if $D<0$.)

\item {\bf Case II:} $D > 0$. Now $k_1^+$ and $k_2^+$ are
  real. This, along with \eq{eq:r12sign} implies that $r_1 r_2
  <0$. Integrating \eq{eq:I13} over $k^+$ yields
\begin{align}
    \label{eq:I15}
    I_1 \Big|_{D > 0} = \frac{\mbox{Sign} (r_1)}{2 \sqrt{D}} .
  \end{align}
  Employing \eq{eq:r12a} we get Sign$(r_1)$ = Sign$[k_1^+ \, (q^+ -
  k_1^+)]$. The $D>0$ case can be realized in the following two ways:
\begin{align}\label{eq:real}
  (a)~(k_T + p_T)^2-2 q^+ q^-<0\qquad\text{or}\qquad(b)~(k_T-p_T)^2-2
  q^+ q^- >0.
\end{align}
For the case (a), after some algebra involving Eqs.~\eqref{eq:k12} and
\eqref{eq:D} one can show that Sign$(r_1)=+1$. For the case (b), one
can similarly show that Sign$(r_1)=-1$. We conclude that
\begin{align}
    \label{eq:I16}
    I_1 \Big|_{D > 0} = \frac{\theta \left(2 q^+ q^- - (k_T + p_T)^2
      \right) - \theta \left( (k_T-p_T)^2-2 q^+ q^- \right)}{2
      \sqrt{D}} .
  \end{align}
\end{itemize}

By combining the above cases I and II, we have
\begin{align}\label{eq:I17}
  I_1 = \frac{\theta\left( 2 q^+ q^- - (k_T+p_T)^2\right) - \theta
    \left( (k_T-p_T)^2-2 q^+ q^- \right) - i\theta(-D) \,
    \mbox{Sign}~(q^\pm)}{2\sqrt{|D|}}.
\end{align}
In order to perform integrations using the complex plane, it is
desirable to rewrite $I_1$ without $\theta$-functions, since the
latter are hard to analytically continue into the whole complex
plane. This is achieved by re-writing \eq{eq:I17} as
\begin{align}
  I_1 & =\frac{1}{4 (q^- + i \epsilon) \left[ q^+ + i \epsilon - \frac{(k_T+p_T)^2}{2( q^- + i \epsilon)} \right]^\frac{1}{2} \ \left[ q^+ + i \epsilon - \frac{(k_T-p_T)^2}{2 (q^- + i \epsilon)} \right]^{\frac{1}{2}} } \notag \\
  & = \frac{1}{4 (q^++ i \epsilon) \left[q^- + i \epsilon -
      \frac{(k_T+p_T)^2}{2 (q^+ + i \epsilon)} \right]^{\frac{1}{2}} \
    \left[ q^- - \frac{(k_T-p_T)^2}{2 ( q^+ + i \epsilon) }
    \right]^{\frac{1}{2}} }.
\end{align}
The $i \epsilon$ regulators are needed for the ``standard'' branch cut
of the square root running along the negative real axis. If we choose
the branch cut of the square root to run along the negative imaginary
axis, one can neglect these $i \epsilon$'s, arriving at the
\eq{eq:I1final} in the main text.



\begin{thebibliography}{10}

\bibitem{Maldacena:1997re}
J.~M. Maldacena, \emph{{The large {N} limit of superconformal field theories
  and supergravity}}, {\emph{Adv. Theor. Math. Phys.} {\bfseries 2} (1998)
  231--252}, [\href{https://arxiv.org/abs/hep-th/9711200}{{\ttfamily
  hep-th/9711200}}].

\bibitem{Witten:1998qj}
E.~Witten, \emph{{Anti-de Sitter space and holography}}, {\emph{Adv. Theor.
  Math. Phys.} {\bfseries 2} (1998) 253--291},
  [\href{https://arxiv.org/abs/hep-th/9802150}{{\ttfamily hep-th/9802150}}].

\bibitem{Gubser:2008pc}
S.~S. Gubser, S.~S. Pufu and A.~Yarom, \emph{{Entropy production in collisions
  of gravitational shock waves and of heavy ions}},
  \href{https://doi.org/10.1103/PhysRevD.78.066014}{\emph{Phys. Rev.}
  {\bfseries D78} (2008) 066014},
  [\href{https://arxiv.org/abs/0805.1551}{{\ttfamily 0805.1551}}].

\bibitem{Lin:2009pn}
S.~Lin and E.~Shuryak, \emph{{Grazing Collisions of Gravitational Shock Waves
  and Entropy Production in Heavy Ion Collision}},
  \href{https://doi.org/10.1103/PhysRevD.79.124015}{\emph{Phys. Rev.}
  {\bfseries D79} (2009) 124015},
  [\href{https://arxiv.org/abs/0902.1508}{{\ttfamily 0902.1508}}].

\bibitem{Kovchegov:2009du}
Y.~V. Kovchegov and S.~Lin, \emph{{Toward Thermalization in Heavy Ion
  Collisions at Strong Coupling}},
  \href{https://doi.org/10.1007/JHEP03(2010)057}{\emph{JHEP} {\bfseries 1003}
  (2010) 057}, [\href{https://arxiv.org/abs/0911.4707}{{\ttfamily 0911.4707}}].

\bibitem{Chesler:2009cy}
P.~M. Chesler and L.~G. Yaffe, \emph{{Boost invariant flow, black hole
  formation, and far-from-equilibrium dynamics in $N = 4$ supersymmetric
  Yang-Mills theory}},
  \href{https://doi.org/10.1103/PhysRevD.82.026006}{\emph{Phys.Rev.} {\bfseries
  D82} (2010) 026006}, [\href{https://arxiv.org/abs/0906.4426}{{\ttfamily
  0906.4426}}].

\bibitem{Chesler:2010bi}
P.~M. Chesler and L.~G. Yaffe, \emph{{Holography and colliding gravitational
  shock waves in asymptotically AdS$_5$ spacetime}},
  \href{https://doi.org/10.1103/PhysRevLett.106.021601}{\emph{Phys.Rev.Lett.}
  {\bfseries 106} (2011) 021601},
  [\href{https://arxiv.org/abs/1011.3562}{{\ttfamily 1011.3562}}].

\bibitem{Baier:2000sb}
R.~Baier, A.~H. Mueller, D.~Schiff and D.~T. Son, \emph{{'Bottom up'
  thermalization in heavy ion collisions}},
  \href{https://doi.org/10.1016/S0370-2693(01)00191-5}{\emph{Phys. Lett.}
  {\bfseries B502} (2001) 51--58},
  [\href{https://arxiv.org/abs/hep-ph/0009237}{{\ttfamily hep-ph/0009237}}].

\bibitem{McLerran:1993ni}
L.~D. McLerran and R.~Venugopalan, \emph{Computing quark and gluon distribution
  functions for very large nuclei}, {\emph{Phys. Rev.} {\bfseries D49} (1994)
  2233--2241}, [\href{https://arxiv.org/abs/hep-ph/9309289}{{\ttfamily
  hep-ph/9309289}}].

\bibitem{McLerran:1993ka}
L.~D. McLerran and R.~Venugopalan, \emph{Gluon distribution functions for very
  large nuclei at small transverse momentum}, {\emph{Phys. Rev.} {\bfseries
  D49} (1994) 3352--3355},
  [\href{https://arxiv.org/abs/hep-ph/9311205}{{\ttfamily hep-ph/9311205}}].

\bibitem{McLerran:1994vd}
L.~D. McLerran and R.~Venugopalan, \emph{Green's functions in the color field
  of a large nucleus}, {\emph{Phys. Rev.} {\bfseries D50} (1994) 2225--2233},
  [\href{https://arxiv.org/abs/hep-ph/9402335}{{\ttfamily hep-ph/9402335}}].

\bibitem{Krasnitz:1998ns}
A.~Krasnitz and R.~Venugopalan, \emph{Non-perturbative computation of gluon
  mini-jet production in nuclear collisions at very high energies},
  {\emph{Nucl. Phys.} {\bfseries B557} (1999) 237},
  [\href{https://arxiv.org/abs/hep-ph/9809433}{{\ttfamily hep-ph/9809433}}].

\bibitem{Krasnitz:1999wc}
A.~Krasnitz and R.~Venugopalan, \emph{The initial energy density of gluons
  produced in very high energy nuclear collisions}, {\emph{Phys. Rev. Lett.}
  {\bfseries 84} (2000) 4309--4312},
  [\href{https://arxiv.org/abs/hep-ph/9909203}{{\ttfamily hep-ph/9909203}}].

\bibitem{Krasnitz:2003nv}
A.~Krasnitz, Y.~Nara and R.~Venugopalan, \emph{Probing a color glass condensate
  in high energy heavy ion collisions}, {\emph{Braz. J. Phys.} {\bfseries 33}
  (2003) 223--230}.

\bibitem{Lappi:2003bi}
T.~Lappi, \emph{Production of gluons in the classical field model for heavy ion
  collisions}, {\emph{Phys. Rev.} {\bfseries C67} (2003) 054903},
  [\href{https://arxiv.org/abs/hep-ph/0303076}{{\ttfamily hep-ph/0303076}}].

\bibitem{Arnold:2003rq}
P.~B. Arnold, J.~Lenaghan and G.~D. Moore, \emph{{QCD plasma instabilities and
  bottom up thermalization}},
  \href{https://doi.org/10.1088/1126-6708/2003/08/002}{\emph{JHEP} {\bfseries
  08} (2003) 002}, [\href{https://arxiv.org/abs/hep-ph/0307325}{{\ttfamily
  hep-ph/0307325}}].

\bibitem{Arnold:2004ih}
P.~Arnold and J.~Lenaghan, \emph{{The abelianization of QCD plasma
  instabilities}}, {\emph{Phys. Rev.} {\bfseries D70} (2004) 114007},
  [\href{https://arxiv.org/abs/hep-ph/0408052}{{\ttfamily hep-ph/0408052}}].

\bibitem{Arnold:2004ti}
P.~Arnold, J.~Lenaghan, G.~D. Moore and L.~G. Yaffe, \emph{Apparent
  thermalization due to plasma instabilities in quark gluon plasma},
  {\emph{Phys. Rev. Lett.} {\bfseries 94} (2005) 072302},
  [\href{https://arxiv.org/abs/nucl-th/0409068}{{\ttfamily nucl-th/0409068}}].

\bibitem{Rebhan:2004ur}
A.~Rebhan, P.~Romatschke and M.~Strickland, \emph{{Hard-loop dynamics of
  non-Abelian plasma instabilities}}, {\emph{Phys. Rev. Lett.} {\bfseries 94}
  (2005) 102303}, [\href{https://arxiv.org/abs/hep-ph/0412016}{{\ttfamily
  hep-ph/0412016}}].

\bibitem{Romatschke:2006nk}
P.~Romatschke and R.~Venugopalan, \emph{The unstable glasma}, {\emph{Phys.
  Rev.} {\bfseries D74} (2006) 045011},
  [\href{https://arxiv.org/abs/hep-ph/0605045}{{\ttfamily hep-ph/0605045}}].

\bibitem{Berges:2013fga}
J.~Berges, K.~Boguslavski, S.~Schlichting and R.~Venugopalan, \emph{{Universal
  attractor in a highly occupied non-Abelian plasma}},
  \href{https://doi.org/10.1103/PhysRevD.89.114007}{\emph{Phys. Rev.}
  {\bfseries D89} (2014) 114007},
  [\href{https://arxiv.org/abs/1311.3005}{{\ttfamily 1311.3005}}].

\bibitem{Gelis:2013rba}
T.~Epelbaum and F.~Gelis, \emph{{Pressure isotropization in high energy heavy
  ion collisions}},
  \href{https://doi.org/10.1103/PhysRevLett.111.232301}{\emph{Phys. Rev. Lett.}
  {\bfseries 111} (2013) 232301},
  [\href{https://arxiv.org/abs/1307.2214}{{\ttfamily 1307.2214}}].

\bibitem{Epelbaum:2014yja}
T.~Epelbaum, F.~Gelis and B.~Wu, \emph{{Nonrenormalizability of the classical
  statistical approximation}},
  \href{https://doi.org/10.1103/PhysRevD.90.065029}{\emph{Phys. Rev.}
  {\bfseries D90} (2014) 065029},
  [\href{https://arxiv.org/abs/1402.0115}{{\ttfamily 1402.0115}}].

\bibitem{Epelbaum:2014mfa}
T.~Epelbaum, F.~Gelis, N.~Tanji and B.~Wu, \emph{{Properties of the Boltzmann
  equation in the classical approximation}},
  \href{https://doi.org/10.1103/PhysRevD.90.125032}{\emph{Phys. Rev.}
  {\bfseries D90} (2014) 125032},
  [\href{https://arxiv.org/abs/1409.0701}{{\ttfamily 1409.0701}}].

\bibitem{Mueller:2002gd}
A.~H. Mueller and D.~T. Son, \emph{{On the Equivalence between the Boltzmann
  equation and classical field theory at large occupation numbers}},
  \href{https://doi.org/10.1016/j.physletb.2003.12.047}{\emph{Phys. Lett.}
  {\bfseries B582} (2004) 279--287},
  [\href{https://arxiv.org/abs/hep-ph/0212198}{{\ttfamily hep-ph/0212198}}].

\bibitem{Arnold:2002zm}
P.~B. Arnold, G.~D. Moore and L.~G. Yaffe, \emph{{Effective kinetic theory for
  high temperature gauge theories}},
  \href{https://doi.org/10.1088/1126-6708/2003/01/030}{\emph{JHEP} {\bfseries
  01} (2003) 030}, [\href{https://arxiv.org/abs/hep-ph/0209353}{{\ttfamily
  hep-ph/0209353}}].

\bibitem{Kurkela:2015qoa}
A.~Kurkela and Y.~Zhu, \emph{{Isotropization and hydrodynamization in weakly
  coupled heavy-ion collisions}},
  \href{https://doi.org/10.1103/PhysRevLett.115.182301}{\emph{Phys. Rev. Lett.}
  {\bfseries 115} (2015) 182301},
  [\href{https://arxiv.org/abs/1506.06647}{{\ttfamily 1506.06647}}].

\bibitem{Gribov:1984tu}
L.~V. Gribov, E.~M. Levin and M.~G. Ryskin, \emph{{Semihard Processes in QCD}},
  {\emph{Phys. Rept.} {\bfseries 100} (1983) 1--150}.

\bibitem{Iancu:2003xm}
E.~Iancu and R.~Venugopalan, \emph{The color glass condensate and high energy
  scattering in {QCD}},  \href{https://arxiv.org/abs/hep-ph/0303204}{{\ttfamily
  hep-ph/0303204}}.

\bibitem{Weigert:2005us}
H.~Weigert, \emph{Evolution at small {$x_{bj}$: The Color Glass Condensate}},
  {\emph{Prog. Part. Nucl. Phys.} {\bfseries 55} (2005) 461--565},
  [\href{https://arxiv.org/abs/hep-ph/0501087}{{\ttfamily hep-ph/0501087}}].

\bibitem{Jalilian-Marian:2005jf}
J.~Jalilian-Marian and Y.~V. Kovchegov, \emph{Saturation physics and deuteron
  gold collisions at {RHIC}}, {\emph{Prog. Part. Nucl. Phys.} {\bfseries 56}
  (2006) 104--231}, [\href{https://arxiv.org/abs/hep-ph/0505052}{{\ttfamily
  hep-ph/0505052}}].

\bibitem{Gelis:2010nm}
F.~Gelis, E.~Iancu, J.~Jalilian-Marian and R.~Venugopalan, \emph{{The Color
  Glass Condensate}},
  \href{https://doi.org/10.1146/annurev.nucl.010909.083629}{\emph{Ann.Rev.Nucl.Part.Sci.}
  {\bfseries 60} (2010) 463--489},
  [\href{https://arxiv.org/abs/1002.0333}{{\ttfamily 1002.0333}}].

\bibitem{Albacete:2014fwa}
J.~L. Albacete and C.~Marquet, \emph{{Gluon saturation and initial conditions
  for relativistic heavy ion collisions}},
  \href{https://doi.org/10.1016/j.ppnp.2014.01.004}{\emph{Prog.Part.Nucl.Phys.}
  {\bfseries 76} (2014) 1--42},
  [\href{https://arxiv.org/abs/1401.4866}{{\ttfamily 1401.4866}}].

\bibitem{KovchegovLevin}
Y.~V. Kovchegov and E.~Levin, \emph{Quantum Chromodynamics at High Energy}.
\newblock Cambridge University Press, 2012.

\bibitem{Mueller:1989st}
A.~H. Mueller, \emph{{Small x Behavior and Parton Saturation: A QCD Model}},
  {\emph{Nucl. Phys.} {\bfseries B335} (1990) 115}.

\bibitem{Balitsky:1995ub}
I.~Balitsky, \emph{{Operator expansion for high-energy scattering}},
  \href{https://doi.org/10.1016/0550-3213(95)00638-9}{\emph{Nucl. Phys.}
  {\bfseries B463} (1996) 99--160},
  [\href{https://arxiv.org/abs/hep-ph/9509348}{{\ttfamily hep-ph/9509348}}].

\bibitem{Balitsky:1998ya}
I.~Balitsky, \emph{Factorization and high-energy effective action},
  {\emph{Phys. Rev.} {\bfseries D60} (1999) 014020},
  [\href{https://arxiv.org/abs/hep-ph/9812311}{{\ttfamily hep-ph/9812311}}].

\bibitem{Kovchegov:1999yj}
Y.~V. Kovchegov, \emph{Small-x {$F_2$} structure function of a nucleus
  including multiple pomeron exchanges}, {\emph{Phys. Rev.} {\bfseries D60}
  (1999) 034008}, [\href{https://arxiv.org/abs/hep-ph/9901281}{{\ttfamily
  hep-ph/9901281}}].

\bibitem{Kovchegov:1999ua}
Y.~V. Kovchegov, \emph{Unitarization of the {BFKL} pomeron on a nucleus},
  {\emph{Phys. Rev.} {\bfseries D61} (2000) 074018},
  [\href{https://arxiv.org/abs/hep-ph/9905214}{{\ttfamily hep-ph/9905214}}].

\bibitem{Jalilian-Marian:1997dw}
J.~Jalilian-Marian, A.~Kovner and H.~Weigert, \emph{The {Wilson}
  renormalization group for low x physics: Gluon evolution at finite parton
  density}, {\emph{Phys. Rev.} {\bfseries D59} (1998) 014015},
  [\href{https://arxiv.org/abs/hep-ph/9709432}{{\ttfamily hep-ph/9709432}}].

\bibitem{Jalilian-Marian:1997gr}
J.~Jalilian-Marian, A.~Kovner, A.~Leonidov and H.~Weigert, \emph{The {Wilson}
  renormalization group for low x physics: Towards the high density regime},
  {\emph{Phys. Rev.} {\bfseries D59} (1998) 014014},
  [\href{https://arxiv.org/abs/hep-ph/9706377}{{\ttfamily hep-ph/9706377}}].

\bibitem{Iancu:2001ad}
E.~Iancu, A.~Leonidov and L.~D. McLerran, \emph{{The renormalization group
  equation for the color glass condensate}},
  \href{https://doi.org/10.1016/S0370-2693(01)00524-X}{\emph{Phys. Lett.}
  {\bfseries B510} (2001) 133--144}.

\bibitem{Iancu:2000hn}
E.~Iancu, A.~Leonidov and L.~D. McLerran, \emph{Nonlinear gluon evolution in
  the color glass condensate. {I}}, {\emph{Nucl. Phys.} {\bfseries A692} (2001)
  583--645}, [\href{https://arxiv.org/abs/hep-ph/0011241}{{\ttfamily
  hep-ph/0011241}}].

\bibitem{Balitsky:2004rr}
I.~Balitsky, \emph{{Scattering of shock waves in QCD}},
  \href{https://doi.org/10.1103/PhysRevD.70.114030}{\emph{Phys. Rev.}
  {\bfseries D70} (2004) 114030},
  [\href{https://arxiv.org/abs/hep-ph/0409314}{{\ttfamily hep-ph/0409314}}].

\bibitem{Chirilli:2015tea}
G.~A. Chirilli, Y.~V. Kovchegov and D.~E. Wertepny, \emph{{Classical Gluon
  Production Amplitude for Nucleus-Nucleus Collisions: First Saturation
  Correction in the Projectile}},
  \href{https://doi.org/10.1007/JHEP03(2015)015}{\emph{JHEP} {\bfseries 03}
  (2015) 015}, [\href{https://arxiv.org/abs/1501.03106}{{\ttfamily
  1501.03106}}].

\bibitem{Kovner:1995ts}
A.~Kovner, L.~D. McLerran and H.~Weigert, \emph{Gluon production at high
  transverse momentum in the mclerran-venugopalan model of nuclear structure
  functions}, {\emph{Phys. Rev.} {\bfseries D52} (1995) 3809--3814},
  [\href{https://arxiv.org/abs/hep-ph/9505320}{{\ttfamily hep-ph/9505320}}].

\bibitem{Kovner:1995ja}
A.~Kovner, L.~D. McLerran and H.~Weigert, \emph{Gluon production from
  non{A}belian {W}eizsacker-{W}illiams fields in nucleus-nucleus collisions},
  {\emph{Phys. Rev.} {\bfseries D52} (1995) 6231--6237},
  [\href{https://arxiv.org/abs/hep-ph/9502289}{{\ttfamily hep-ph/9502289}}].

\bibitem{Kovchegov:1997ke}
Y.~V. Kovchegov and D.~H. Rischke, \emph{Classical gluon radiation in
  ultrarelativistic nucleus nucleus collisions}, {\emph{Phys. Rev.} {\bfseries
  C56} (1997) 1084--1094},
  [\href{https://arxiv.org/abs/hep-ph/9704201}{{\ttfamily hep-ph/9704201}}].

\bibitem{Kovchegov:1998bi}
Y.~V. Kovchegov and A.~H. Mueller, \emph{Gluon production in current nucleus
  and nucleon nucleus collisions in a quasi-classical approximation},
  {\emph{Nucl. Phys.} {\bfseries B529} (1998) 451--479},
  [\href{https://arxiv.org/abs/hep-ph/9802440}{{\ttfamily hep-ph/9802440}}].

\bibitem{Dumitru:2001ux}
A.~Dumitru and L.~D. McLerran, \emph{How protons shatter colored glass},
  {\emph{Nucl. Phys.} {\bfseries A700} (2002) 492--508},
  [\href{https://arxiv.org/abs/hep-ph/0105268}{{\ttfamily hep-ph/0105268}}].

\bibitem{Gelis:2008rw}
F.~Gelis, T.~Lappi and R.~Venugopalan, \emph{{High energy factorization in
  nucleus-nucleus collisions}},
  \href{https://doi.org/10.1103/PhysRevD.78.054019}{\emph{Phys.Rev.} {\bfseries
  D78} (2008) 054019}, [\href{https://arxiv.org/abs/0804.2630}{{\ttfamily
  0804.2630}}].

\bibitem{Gelis:2008ad}
F.~Gelis, T.~Lappi and R.~Venugopalan, \emph{{High energy factorization in
  nucleus-nucleus collisions. II. Multigluon correlations}},
  \href{https://doi.org/10.1103/PhysRevD.78.054020}{\emph{Phys.Rev.} {\bfseries
  D78} (2008) 054020}, [\href{https://arxiv.org/abs/0807.1306}{{\ttfamily
  0807.1306}}].

\bibitem{Kovchegov:2005ss}
Y.~V. Kovchegov, \emph{{Can thermalization in heavy ion collisions be described
  by QCD diagrams?}}, {\emph{Nucl. Phys.} {\bfseries A762} (2005) 298--325},
  [\href{https://arxiv.org/abs/hep-ph/0503038}{{\ttfamily hep-ph/0503038}}].

\bibitem{Kovchegov:2005kn}
Y.~V. Kovchegov, \emph{Thoughts on non-perturbative thermalization and jet
  quenching in heavy ion collisions}, {\emph{Nucl. Phys.} {\bfseries A764}
  (2006) 476--497}, [\href{https://arxiv.org/abs/hep-ph/0507134}{{\ttfamily
  hep-ph/0507134}}].

\bibitem{WuKovchegov}
B.~Wu and Y.~V. Kovchegov, ``{Time-Dependent Observables in Heavy Ion
  Collisions I: Setting up the Formalism}.'' in preparation, 2017.

\bibitem{Gelis:2006yv}
F.~Gelis and R.~Venugopalan, \emph{{Particle production in field theories
  coupled to strong external sources}},
  \href{https://doi.org/10.1016/j.nuclphysa.2006.07.020}{\emph{Nucl. Phys.}
  {\bfseries A776} (2006) 135--171},
  [\href{https://arxiv.org/abs/hep-ph/0601209}{{\ttfamily hep-ph/0601209}}].

\bibitem{Lappi:2006hq}
T.~Lappi, \emph{{Energy density of the glasma}},
  \href{https://doi.org/10.1016/j.physletb.2006.10.017}{\emph{Phys. Lett.}
  {\bfseries B643} (2006) 11--16},
  [\href{https://arxiv.org/abs/hep-ph/0606207}{{\ttfamily hep-ph/0606207}}].

\bibitem{Epelbaum:2015vxa}
T.~Epelbaum, F.~Gelis, S.~Jeon, G.~Moore and B.~Wu, \emph{{Kinetic theory of a
  longitudinally expanding system of scalar particles}},
  \href{https://doi.org/10.1007/JHEP09(2015)117}{\emph{JHEP} {\bfseries 09}
  (2015) 117}, [\href{https://arxiv.org/abs/1506.05580}{{\ttfamily
  1506.05580}}].

\bibitem{Lepage:1980fj}
G.~P. Lepage and S.~J. Brodsky, \emph{Exclusive processes in perturbative
  quantum chromodynamics}, {\emph{Phys. Rev.} {\bfseries D22} (1980) 2157}.

\bibitem{Cutkosky:1960sp}
R.~Cutkosky, \emph{{Singularities and discontinuities of Feynman amplitudes}},
  {\emph{J.Math.Phys.} {\bfseries 1} (1960) 429--433}.

\end{thebibliography}


\providecommand{\href}[2]{#2}\begingroup\raggedright\endgroup


\end{document}